 \documentclass[twocolumn,twocolappendix]{aastex631}

\usepackage{graphicx}	% Including figure files
\usepackage{amsmath}	% Advanced maths commands
\usepackage{amssymb}	% Extra maths symbols
\usepackage{float}
\usepackage{mathtools}
\usepackage{xcolor}
\makeatletter
\newcommand*{\hyperlinkcite}[1]{\hyper@link{cite}{cite.#1}}%\hyperlinkcite takes 2 arguments: #1<- cite-key, #2<- link-text
\newcommand{\OmK}{\Omega_\mathrm{K}}

\makeatother

\newcommand{\PaperI}{\hyperlinkcite{lehmann2024}{Paper I}}

\shorttitle{Convective Overstability in Protoplanetary Disks}
\shortauthors{M. Lehmann, M.-K. Lin}

\allowdisplaybreaks

\begin{document}

\title{Convective Overstability in Radially Global Protoplanetary Disks. II. Impact on planetesimal formation}

\author[0000-0002-0496-3539]{Marius Lehmann}
\email{mlehmann@asiaa.sinica.edu.tw}
\affiliation{Institute of Astronomy and Astrophysics, Academia Sinica, Taipei 10617, Taiwan}

\author[0000-0002-8597-4386]{Min-Kai Lin}
\affiliation{Institute of Astronomy and Astrophysics, Academia Sinica, Taipei 10617, Taiwan}
\affiliation{Physics Division, National Center for Theoretical Sciences, Taipei 10617, Taiwan}

\begin{abstract}

The Convective Overstability (COS) is a hydrodynamic instability occurring in protoplanetary disc (PPD) regions with an adverse radial entropy gradient. It is a potential driver of turbulence and may influence planetesimal formation. In this second paper of our series, we study the effects of the COS on dust dynamics in radially global PPD simulations, focusing on the mid-plane region, where vertical gravity on the dust is included. 

Axisymmetric 2D simulations show susceptibility to both the COS and the Vertically Shearing Streaming Instability. For a Stokes number $\tau=0.1$, strong dust clumping occurs only for highly super-solar initial metallicities $Z \gtrsim 0.05$.

In 3D non-axisymmetric simulations, the COS generates large-scale, long-lived vortices that have the potential to efficiently concentrate dust, with dust accumulation strengthening as $\tau$ increases. For $\tau = 0.01$, no strong clumping occurs even at metallicities as high as $Z = 0.1$, and vortices remain robust and long-lived. At $\tau = 0.04$, strong dust clumping is observed for solar metallicity ($Z = 0.01$) and higher. For $\tau = 0.1$, clumping occurs even at strongly sub-solar metallicities ($Z \gtrsim 0.004$), peaking at $Z \sim 0.01$–$0.03$, including solar values. Under these conditions, vortices weaken significantly and become more spatially extended. At higher metallicities ($Z \gtrsim 0.04$) with $\tau = 0.1$, large-scale vortex formation is suppressed, leading to nearly axisymmetric dust rings, which can still undergo clumping via the classical Streaming Instability.

\end{abstract}


% \section{DISCUSS}

% \begin{itemize}

% \item behaviour of dust scaleheight with increasing $Z_0$, why does it decrease, even in isothermal limit. Vertical buoyancy outweighs vertical shear, but it seems so small?? Correlation times?, Hsu and Lin 2022, Yang 2018.

%     \item vortex supression for $\tau=0.1$ and $Z\geq 0.05$ but not for $\tau=0.01$? Discuss radial profiles $Z(r)$. Robustness of results, test simulations. Reduced diagnostic domain.

% \item increased dust scalehieght in vortices, cases with and without feedback

% \item elliptic instability?

% \end{itemize}

% \section{Discuss}

% \begin{itemize}

%     \item earlier vortex formation near inner boundary with increasing Z? it happens gradually earlier with increasing Z. Even at $Z=0.002$ there is difference to the pure gas case.

%     \item fourier analyse output, axisymmetric angular momentum contrbution? nonlinear interations. complex interpretation.
    
%     \item discuss how large vortices actually form. Different way than in raettig2021?

%     \item KHI shear versus VSSI (raettig page 10)?

%     \item dut scaleheight, eddy times, \citet{youdin2007}.

%     \item components of vorticity vector. $\omega_r$ (KHI) and $\omega_z$.

%     \item zonal flow profiles at inner boundary change width with $Z$ ??

%     \item value of $\kappa^2$ (youdines new paper).

%     \item calculate eddie time for 2D simulations.

%     \item see if zonal flow near inner boundary forms also in 2D simulations! isothermal and non-isothermal. vssi with increasing cos, or cos with increasing vssi.
    
% \end{itemize}
\section{Introduction}\label{sec:intro}

Protoplanetary disks are believed to harbor regions where the Magneto-Rotational Instability (MRI: \citealt{balbus1991}) is inactive, commonly referred to as dead zones \citep{gammie1996,turner2009,armitage2011,turner2014}. In these regions, purely hydrodynamic instabilities, such as the Vertical Shear Instability (VSI:  \citealt{urpin1998,urpin2003,nelson2013,barker2015,lin2015}) and Convective Overstability (COS:  \citealt{klahr2014,lyra2014,latter2016}), can still develop. Although these instabilities are generally too weak to account for the global mass accretion rates in protoplanetary disks (PPDs), they can play a crucial role in influencing dust distribution and dynamics (e.g. \citealt{stoll2016,lin2017, flock2017,flock2020, lehmann2022,lehmann2023,lin2025}. However, the interaction between these instabilities and the behavior of small dust grains—key building blocks in the planetesimal formation process under the core accretion model \citep{safronov1972,drazkowska2022}—remains poorly understood. Current planetesimal formation theories propose that dust grains become concentrated into clumps through the Streaming Instability (SI; \citealt{youdin2005,youdin2007,johansen2007}), which subsequently collapse under self-gravity to form planetesimals \citep{johansen2007,bai2010,simon2016,carrera2021a}. However, for small dust grains, the SI requires metallicities (dust-to-gas ratios) several times the solar value to induce sufficient clumping (\citealt{li2021}). More recently, \citet{lim24} demonstrated that when turbulence at plausible levels ($\alpha\sim 10^{-4} - 10^{-3}$) is included, the required metallicity for clumping increases by up to an order of magnitude. These findings highlight the need for dust-concentrating mechanisms to operate prior to the onset of the SI, in order to sustain the planet formation paradigm via this process.

In \citet{lehmann2024} (\hyperlinkcite{lehmann2024}{Paper I} hereafter), we examined in detail the nonlinear saturation of the COS in a pure gas using high resolution 2D axisymmetric and 3D simulations. A key aspect of that study was the radially global setup of the disk region, which differs from previous local models (see below) by allowing the radial disk structure to evolve freely under prescribed boundary conditions.
A key result of \PaperI~was that COS leads to the formation of long-lived vortices, which in turn drive significant radial angular momentum transport. These vortices were found to migrate radially inward at significant rates of $\sim 0.01$ gas pressure scaleheights per orbit (\hyperlinkcite{lehmann2024}{Paper I}). We also demonstrated that this transport can reshape the disk structure on large scales. In particular, it can result in the formation of pressure bumps. Interestingly, simulations by \citet{lehmann2022} suggest that pressure bumps may serve as favorable sites for large-scale vortex formation. Another significant finding was that COS-induced vortices are generally subject to the elliptic instability, which can limit their lifetime, and which is expected to influence dust settling within such vortices.

In this second paper of the series, we build on these findings to investigate the impact of COS on dust dynamics. Specifically, we examine how COS-induced zonal flows and vortices influence dust concentration and whether dust can remain trapped in these structures despite the expected vertical stirring from COS and the elliptic instability. Additionally, dust-gas drag instabilities—such as the SI or the vertically shearing streaming instability (VSSI: \citealt{lin2021})—are expected to be present and influence dust evolution throughout the disk. These instabilities are well-established in non-vortical regions and are known to play a crucial role in dust dynamics. Furthermore, dust-induced instabilities have also been observed within vortices in previous studies \citep{fu2014, rubsamen2015, raettig2015, magnan2024}, though their exact nature remains uncertain. In particular, it is currently unclear whether the SI or the VSSI can actively operate within vortices, though ongoing research suggests this possibility (N. Magnan, private communication). 
% \mkl{probably more appropriate to cite it as private communication}

Compared to previous studies, our work extends the understanding of dust dynamics in the presence of COS in several key ways.
Previous local shearing box simulations by \citet{lyra2018} and \citet{raettig2021} demonstrated that COS can effectively concentrate dust grains within vortices, potentially reaching densities that exceed the Roche limit—a critical condition for self-gravitational collapse and planetesimal formation. \citet{lyra2024} included self-gravity, allowing them to simulate the formation of planetesimals. However, in their simulations, dust grains—represented by Lagrangian superparticles—were introduced only after COS had saturated, enabling their evolution to be tracked over several tens of orbits. In contrast, this work examines the simultaneous evolution of dust and gas starting from equilibrium by employing radially global two-fluid simulations (2D axisymmetric and 3D). This approach accounts for the dust’s influence on COS during its linear growth phase \citep{lehmann2023} and subsequent nonlinear saturation, including the formation and radial migration of COS-induced vortices. Furthermore, it allows for a self-consistent saturation of the instability in response to the radial evolution of both dust and gas densities, which results from the angular momentum transport driven by vortices, as well as radial dust drift in response to gas pressure variations. 

Recently, \citet{lin2025} studied the nonlinear saturation of COS in the presence of dust using local, unstratified, high-resolution axisymmetric 2D simulations in the Boussinesq approximation. They showed that zonal flows induced by COS can concentrate dust with enhancement factors of the order of 10. However, they also found that dust feedback tends to suppress the formation of zonal flows—the precursors to vortices—even for small dust-to-gas ratios of the order of 0.1.

Following \citet{raettig2021}, and in contrast to \citet{lin2025}, we also include the effect of vertical gravity on the dust. As we will see below, this leads to the emergence of the vertically shearing streaming instability (VSSI), discovered by \citet{lin2021}, which can coexist with COS.

These advancements provide a more comprehensive and realistic picture of how COS shapes dust dynamics in protoplanetary disks.

Finally, while 3D simulations offer a more comprehensive physical picture, we also include 2D axisymmetric runs in our study for several key reasons. First, many previous 2D studies have been highly informative for the Convective Overstability (COS)—notably \citet{klahr2014}, \citet{teed2021}, \hyperlinkcite{lehmann2024}{Paper I}, and \citet{lin2025}, as well as 2D theoretical analyses \citep{lehmann2023}. By incorporating 2D simulations, we can directly link our findings to this existing body of work. Second, 2D setups remain widely used for other disk instabilities, largely due to their lower computational costs \citep[e.g., recent  VSI studies by][]{svanberg2022,fukuhara2023,pfeil2024,yun2025b} and the viability of axisymmetric theoretical treatments (e.g. \citealt{yun2025a, ogilvie2025}). We therefore expect that 2D simulations will continue to play an important role also in the case of the COS. Finally, 2D simulations provide a practical first step to gain insights into dust-gas coupling under COS before introducing the additional complexities of full three-dimensional effects and supplementary instabilities (e.g., elliptic \citep{lesur2009} or Rossby wave \citep{lovelace1999} instabilities). This staged approach clarifies the core physical mechanisms and facilitates a direct comparison with both past and future 2D studies.

The paper is structured as follows: Section \ref{sec:model} introduces our hydrodynamic model of a protoplanetary disk. In Section \ref{sec:instabilities}, we briefly discuss potential instabilities in our simulations. Section \ref{sec:hydrosim} details the setup of our hydrodynamic simulations, including the numerical methods, boundary conditions, and key diagnostics used to analyze the results. Sections \ref{sec:sim_2d} and \ref{sec:sim_3d} present the results of our 2D and 3D simulations, respectively. Finally, Section \ref{sec:discussion} summarizes our findings, discusses limitations, and outlines prospects for future research.

\section{Hydrodynamic Model}\label{sec:model}

\subsection{Basic Equations}\label{sec:eqn}

We consider a  global hydrodynamic model of a PPD consisting of gas and a single species of dust, governed by the set of dynamical equations:
\begin{align}
\left(\frac{\partial}{\partial t} + \boldsymbol{v}_{g}\cdot\boldsymbol{\nabla}\right) \, \rho_{g} &=  - \rho_{g} \left( \boldsymbol{\nabla} \cdot \boldsymbol{v}_{g} \right), \label{eq:contrhog}
\\
\begin{split}
 \left(\frac{\partial}{\partial t} + \boldsymbol{v}_{g}\cdot\boldsymbol{\nabla}\right) \, \boldsymbol{v}_{g} &= - \frac{1}{\rho_{g}} \boldsymbol{\nabla} P +  \frac{1}{\rho_{g}} \boldsymbol{\nabla} \cdot \boldsymbol{S} - \boldsymbol{\nabla} \Phi_* \\
 &\quad - \frac{\epsilon}{t_{s}} \left(\boldsymbol{v}_{g} -\boldsymbol{v}_d \right),
\end{split} \label{eq:contvg}
\\
\left(\frac{\partial}{\partial t} + \boldsymbol{v}_{g}\cdot\boldsymbol{\nabla}\right) \, P &=  - \gamma P \left( \boldsymbol{\nabla} \cdot \boldsymbol{v}_{g} \right) - \Lambda, \label{eq:contp}
\\
\begin{split}
\left(\frac{\partial}{\partial t} + \boldsymbol{v}_{d}\cdot\boldsymbol{\nabla}\right) \, \rho_{d} &=  - \rho_{d} \left( \boldsymbol{\nabla} \cdot \boldsymbol{v_{d}} \right) \\
&\quad + \boldsymbol{\nabla} \cdot \left( \nu \rho_g \boldsymbol{\nabla} \left(\frac{\rho_d}{\rho_g} \right) \right),
\end{split} \label{eq:contrhod}
\\
\left(\frac{\partial}{\partial t} + \boldsymbol{v}_{d}\cdot\boldsymbol{\nabla}\right) \, \boldsymbol{v}_{d} &= - \boldsymbol{\nabla}\Phi_* - \frac{1}{t_{s}} \left(\boldsymbol{v}_{d} - \boldsymbol{v}_{g} \right), \label{eq:contvd}
\end{align}
In these equations $\rho_{g}$, $\rho_{d}$, $\boldsymbol{v}_{g}$, $\boldsymbol{v}_{d}$ and $P$ represent the gas and dust volume mass densities, the three-dimensional gas and dust velocities, and the gas pressure, respectively. The dust-to-gas density ratio is given by:
\begin{align}
\epsilon & = \frac{\rho_{d}}{\rho_{g}}\label{eq:eps}.
\end{align}
In addition, we define the metallicity as the ratio of the dust and gas surface mass densities:
\begin{equation}
    Z = \frac{\Sigma_d}{\Sigma_g}.
\end{equation}
This quantity represents the vertically integrated dust-to-gas mass ratio. While $Z$ can vary radially, it does not capture variations in the vertical direction. By contrast $\epsilon$ varies both radially and vertically and characterizes the dust concentration at a given location in the disk.

We adopt a non-rotating frame with cylindrical coordinates $(r,\varphi,z)$ and with origin on a central star of mass $M_{*}$ with gravitational potential $\Phi_{*}$.
Self-gravity, magnetic fields, and the indirect gravitational term are neglected.

The gas pressure is assumed to follow an ideal gas equation of state:
\begin{equation}\label{eq:eos}
P=  \frac{\mathcal{R}}{\mu} \rho_{g} T
\end{equation}
where $\mathcal{R}$ is the gas constant, $\mu$ is the mean molecular weight, and $T$ is the gas temperature. We also define the effective gas sound speed as
\begin{equation}
    c_s \equiv \sqrt{\frac{P}{\rho_g}}.
\end{equation}

Following \citet{klahr2014}, we assume that gas cooling occurs in the optically thin regime, with the cooling term defined as:
\begin{equation}
    \Lambda  = \frac{\rho_{g} \mathcal{R}}{\mu}\frac{\delta T}{t_{c}},
\end{equation}
where $\delta T$ represents the deviation of the temperature from its equilibrium profile (defined below). 
%This cooling term also incorporates viscous heating. 
We express the cooling time using the dimensionless parameter:
\begin{equation}
\beta = \OmK(r) t_{c} .
\end{equation}
where $\OmK(r)$ is the local Keplerian angular velocity. The adiabatic index is fixed at $\gamma = 1.4$.

The viscous stress tensor\footnote{ Here, $\dagger$ denotes the conjugate transpose, and $\boldsymbol{I}$ is the identity tensor.} is given by:
\begin{equation}\label{eq:stresstensor}
\boldsymbol{S}= \rho \nu \left( \boldsymbol{\nabla} \boldsymbol{v} + \left( \boldsymbol{\nabla} \boldsymbol{v} \right) ^{\dagger} - \frac{2}{3} \boldsymbol{I} \boldsymbol{\nabla} \cdot \boldsymbol{v} \right)
\end{equation}
where $\nu$ is the kinematic viscosity, which also parameterizes dust diffusion in the last term of Eq.~(\ref{eq:contrhod}).
In the nonlinear simulations presented below, viscosity is included solely to ensure numerical stability. The value of $\nu$ is chosen to be very small, significantly lower than the typical turbulent viscosities  measured in 3D simulations.

\subsection{Dust-gas drag}

The dust component is modeled as a pressureless fluid interacting with the gas through a frictional force \citep{johansen2014}, characterized by the stopping time:

\begin{equation}
t_{s} = \frac{r_{p} \rho_{p}}{\rho_{g} c_{s}},
\end{equation}

where  $r_{p}$  and  $\rho_{p}$  denote the particle radius and bulk density, respectively. The stopping time represents the characteristic timescale over which a dust grain adjusts its velocity to match the surrounding gas. The fluid approximation for dust is valid when  $t_{s}$  is sufficiently small \citep{jacquet2011}.

We typically work with the dimensionless Stokes number, defined as:

\begin{equation}
\tau = \Omega_{K} t_{s},
\end{equation}

where $\tau \ll 1$ indicates that dust grains are strongly, though not perfectly, coupled to the gas.  

Throughout this work, a subscript “0” denotes a reference value at $r = r_0$, $z = 0$, and $t = 0$. Accordingly, the Stokes number at this reference location and time, denoted as $\tau_0$, is a free parameter in our model. For simplicity, we refer to $\tau_0$ as the Stokes number $\tau$ when describing simulation results below.

In our simulations, the stopping time is computed at any given location and time using:

\begin{equation}\label{eq:tstop}
t_{s} = \frac{\tau_{0}}{\Omega_{K}(r_{0})} \frac{\rho_{g0} c_{s0}}{\rho_{g} c_{s}},
\end{equation}

where the gas density $\rho_{g}$ and sound speed $c_{s}$ are evaluated locally and normalized to their initial values at the reference location.

\subsection{Vertical Gravity on the dust}

As in \hyperlinkcite{lehmann2024}{Paper I}, we focus on a region close to the disk mid-plane, where  $|z| \ll H_{g}$, with  $H_{g} = c_s / \Omega_K$  denoting the gas pressure scale height. In this regime, the stellar gravitational potential can be approximated as:

\begin{equation}\label{eq:pot}
\Phi_* = \frac{G M_{}}{\sqrt{r^2 + z^2}} \approx \frac{G M_{}}{r} - \frac{1}{2}\Omega_K^2 z^2,
\end{equation}

where the first term corresponds to the radial component, and the second term captures the leading-order vertical variation.

Since dust can settle into a thin mid-plane layer, we incorporate the vertical variation of  $\Phi_*$  from Eq.(\ref{eq:pot}) in the dust momentum equation, Eq.(\ref{eq:contvd}). However, we neglect the vertical variation of  $\Phi_*$  in the gas momentum equation, Eq.~(\ref{eq:contvg}), because the gas density varies only slightly on the vertical scales considered. This approximation, also employed by \citet{raettig2021}, effectively eliminates the classic Vertical Shear Instability (VSI) from the problem, as the gas itself exhibits no vertical shear. Additionally, it facilitates the use of vertical periodic boundary conditions.
Nevertheless, due to the vertically stratified dust layer, vertical shear $( \partial \Omega / \partial z )$ in the dust-gas mixture is generally non-zero, which may trigger the VSSI.

\subsection{Dust $\&$ gas initial state}\label{sec:disc_equil}

As in \hyperlinkcite{lehmann2024}{Paper I}, we assume an axisymmetric disk equilibrium with radial power-law profiles for the gas volume mass density:
\begin{equation}
\rho_{g}(r) = \rho_{g0} \left(\frac{r}{r_{0}}\right)^{-p},
\end{equation}
and temperature:
\begin{equation}
T(r) = T_{0} \left(\frac{r}{r_{0}}\right)^{-q}.
\end{equation}

Formally, the initial dust density in our simulations is determined by balancing vertical settling, driven by stellar gravity (with a settling velocity of $v_{dz} = -\tau \OmK z$), and initial vertical diffusion \citep[e.g.,][]{lin2021}. The resulting equilibrium profile is given by:
\begin{equation}\label{eq:dustdens}
\rho_d(r, z) = \rho_{d0} \exp\left(-\frac{z^2}{2H_d^2}\right),
\end{equation}
where $\rho_{d0} = \epsilon_0 \rho_{g0}$, and the dust scale height $H_d$ is given by:
\begin{equation}\label{eq:dusthd}
H_d = \sqrt{\frac{\delta}{\tau}} H_g,
\end{equation}
with the dust diffusion coefficient $\delta$ defined as:
\begin{equation}\label{eq:delta}
\delta = \frac{\nu}{c_s H_g}.
\end{equation}
Here, we assume $\tau \ll 1$ and $\delta \ll \tau$. Since the viscosity $\nu$ is set to a very small value, the corresponding dust diffusion coefficient $\delta$ is also small, leading to an equilibrium dust scale height $H_d$ that is unrealistically thin and cannot be resolved. However, as the dust layer collapses, turbulence driven by the VSSI and COS develops in our simulations, significantly increasing the effective value of $\delta$.

Therefore, instead of initializing the dust layer with Eq. (\ref{eq:dusthd}) we follow \citet{li2021} and initially set $H_d = \frac{1}{2} \eta r$, and assume the \citet{nakagawa1986} equilibrium for the dust and gas velocities as functions of height $z$:
\begin{equation}
\begin{split}
    \boldsymbol{v}_{g} & =\left(  \sqrt{1-2 f_g \eta} +f_g^3 \frac{\eta  \tau^2}{1+ (\tau f_g)^2}\right) \OmK r \boldsymbol{e}_{\varphi} \\
    \quad & + 2 f_g f_d \frac{\eta  \tau}{1 + \left(f_g \tau\right)^2}\OmK r   \boldsymbol{e}_r
\end{split}
\end{equation}
and:
\begin{equation}
\begin{split}
    \boldsymbol{v}_{d} & =\left( \sqrt{1-2 f_g \eta} -f_d f_g^2 \frac{\eta  \tau^2}{1+ (\tau f_g)^2}\right) \OmK r  \boldsymbol{e}_{\varphi} \\
    \quad & - f_g^2 \frac{\eta\tau}{1 + \left(f_g \tau\right)^2}\OmK r   \boldsymbol{e}_r,
\end{split}
\end{equation}
where the dimensionless radial pressure gradient is given by \citep{youdin2005}:
\begin{equation}\label{eq:eta}
\eta(r) \equiv -\frac{1}{2 \rho_g (\OmK r )^2} \frac{\partial P}{\partial \ln r} = \frac{h^2}{2}\left(p+q\right),
\end{equation}
with the total density $\rho = \rho_{g} + \rho_{d}$, and the dust and gas fractions:
\begin{equation*}
f_g=\frac{\rho_g}{\rho}, \quad f_d=\frac{\rho_d}{\rho}.
\end{equation*}
Neglecting initial planar velocity deviations caused by dust-gas drag (i.e., terms proportional to $\tau$ or $\tau^2$) is not consequential, as the system rapidly adjusts to the correct equilibrium. As in \citet{li2021}, we neglect vertical dust settling in the initial state; however, dust begins to settle immediately as we evolve Eqs.~(\ref{eq:contrhog})–(\ref{eq:contvd}) in our simulations.

The squared radial buoyancy frequency, which must be negative to drive the COS, is given by:
\begin{equation}\label{eq:nr2}
N_r^2  \equiv -\frac{1}{\gamma\rho_{g}}\frac{\partial P}{\partial r}\frac{\partial S}{\partial r} =  -\frac{1}{\gamma} h^2 \OmK^2 \left(p+q\right)\left(q+\left[1-\gamma\right]p\right),
\end{equation}
where the dimensionless gas entropy is defined as:
\begin{equation}\label{eq:ent}
S \equiv \ln{\frac{P}{\rho_{g}^\gamma}},
\end{equation}
and the disk aspect ratio is given by $h = H_{g}/r$. For a discussion on the disk regions that may support $N_r^2 < 0$, the reader is referred to \PaperI \ and references therein.
The second equality in Eqs.~(\ref{eq:eta}) and (\ref{eq:nr2}) assumes the radial power-law profiles defined above. For further details on the COS mechanism, we refer the reader to \hyperlinkcite{lehmann2024}{Paper I} and references therein.

Since vertical gravity is included for the dust, the dusty gas mixture is, in principle, subject to vertical buoyancy as well. Following \citet{chiang08}, the squared vertical buoyancy frequency is defined as:
\begin{equation}\label{eq:Nz2}
N_{z}^2 \equiv -\OmK^2 z \frac{\partial \rho_{d}}{\partial z},
\end{equation}
where Eq.~(\ref{eq:pot}) is used, and it is assumed that $|\partial \rho_{g}/\partial z| \ll |\partial \rho_{d}/\partial z|$. However, in our simulations, we find that $N_z^2$ typically assumes dynamically insignificant values. 

Another aspect introduced by dust vertical gravity is vertical shear of the dust-gas mixture $r\frac{\partial \Omega}{\partial z}$, where $\Omega = v/r$ with the dust-gas center of mass velocity $v = f_g v_g + f_d v_d$. Vertical shear turns out to be significant, and results in the onset of the VSSI in our simulations, as shown below.

\section{Potential Instabilities}\label{sec:instabilities}

Before presenting our numerical simulations, we first describe and compare the key instabilities that may be relevant in our setup. This overview will aid in interpreting our results.

\subsection{Convective Overstability (COS)}
The COS, the main focus of this study, corresponds to overstable inertial waves in the gas \citep{latter2016}. It is an axisymmetric instability that requires a vertical dimension but does not depend on stratification. The COS operates under conditions of adverse radial buoyancy, i.e., $N_r^2 < 0$, and its growth rates peak when the cooling time is comparable to the dynamical timescale, $\beta \sim 1$ \citep{lyra2014}. In this regime, buoyancy forces remain slightly out of phase with fluid parcel displacements, driving overstable oscillations. The presence of well-coupled dust can weaken the COS by effectively reducing the magnitude of $|N_r^2|$ \citep{lehmann2023}.

\subsection{Subcritical Baroclinic Instability (SBI)}
Beyond the linear COS, protoplanetary disks can also sustain a subcritical baroclinic instability (SBI) \citep[e.g.,][]{klahr2003, lesur2010}. Unlike linear instabilities, SBI does not grow from infinitesimal perturbations but instead requires a finite-amplitude seed vortex. Once a sufficiently strong vortex forms, baroclinic torques can sustain or even amplify its circulation by tapping into misaligned pressure and density gradients. This process relies on retaining local thermal or entropy gradients over the vortex turnover timescale, which is more effective at longer cooling times ($\beta \gtrsim 1$). Conversely, when $\beta \ll 1$, temperature perturbations within the vortex are quickly erased, quenching baroclinic amplification.

Although our simulations do not explicitly test SBI, we note that \PaperI \ reported significantly enhanced vortex lifetimes and strengths at $\beta = 10$ compared to $\beta = 1$, consistent with baroclinic vortex amplification \citep[see also][]{raettig2013}.

\subsection{Streaming Instability (SI)}
The SI arises from the relative drift between dust and gas in a dusty, unstratified disk \citep{youdin2005}. In the dust-poor limit ($f_d \ll 1$), the SI can be understood as a resonance between inertial waves in the gas and the relative drift of dust, leading to an instability that efficiently concentrates dust \citep{squire2018, magnan2024}. This classification identifies the SI as a \emph{resonant drag instability} \citep{squire2018}. Like the COS, the SI is axisymmetric but requires a vertical dimension to operate. Its growth rates peak when both $\tau$ and the local dust-to-gas ratio are of order unity, as the instability relies on dust backreaction onto the gas.  

\subsection{Vertically Shearing Streaming Instability (VSSI)}  
When the dust layer is stratified under vertical gravity, a vertical gradient in rotation velocity develops between the gas-dominated layers above and below the dust-dominated midplane. This gradient drives the VSSI \citep{ishitsu09, lin2021}, a drag instability distinct from the Streaming Instability (SI). While SI is powered by relative motion between dust and gas, VSSI arises from vertical shear within the dust-gas mixture. VSSI is most active near the dust surface, where shear is strongest, operating on smaller length scales with higher growth rates than SI. 

In the theoretical limit of \( \tau \to 0 \), where dust would become perfectly coupled to the gas, VSSI is expected to vanish, leaving a single fluid with vertical shear that could, in principle, become susceptible to the Kelvin-Helmholtz Instability (KHI, e.g., \citealt{johansen2006,chiang08}). However, our simulations show no KHI signatures—such as characteristic roll-up structures or enhanced vorticity in the azimuthal-vertical plane—likely because the significant vertical shear driving VSSI generates turbulence and vorticity patterns that disrupt the coherent shear layers required for KHI onset, despite a potentially low Richardson number (\( \mathrm{Ri} = N_z^2 / (\partial v_\phi / \partial z)^2 \)). This suggests that VSSI, by dominating the dynamics, may suppress KHI within the parameter space explored in this study.

\subsection{Elliptic Instability}  
Vortices commonly emerge from hydrodynamic instabilities, including the COS \citep{lyra2014, lehmann2024}. However, vortices themselves can be susceptible to the Elliptic Instability \citep{lesur2009, railton14}. Vortices with aspect ratios $\chi \lesssim 4$ are particularly vulnerable, as the instability disrupts their coherence by generating small-scale turbulence. In strongly unstable cases, it can lead to complete vortex destruction. While vortices have been proposed as efficient dust traps, turbulence induced by the elliptic instability can diffuse dust grains, leading to a steady-state distribution rather than sustained clumping \citep{lyra2013}.

\leavevmode\bigskip

\section{Numerical Method and diagnostics}\label{sec:hydrosim}

% \subsection{Numerical setup and diagnostics}

\subsection{Simulation setup}

We conduct hydrodynamic simulations to solve equations (\ref{eq:contrhog})---(\ref{eq:contvd}) using FARGO3D\footnote{\url{http://fargo.in2p3.fr}} \citep{fargo3d, llambay2019}. Given that our disk model is vertically unstratified, we apply periodic boundary conditions in the azimuthal and vertical directions for all quantities.
As in \hyperlinkcite{lehmann2024}{Paper I}, the radial boundary conditions extrapolate the equilibrium values of mass density and azimuthal velocity into the ghost zones, while the radial and vertical velocities are set to zero at the boundaries. Additionally, damping zones are applied at the outer boundary to restore all field variables to their equilibrium values. This approach ensures that the initial dust density profile is maintained, providing a continuous supply of dust into the simulation domain. Without this measure, dust would rapidly deplete as it migrates inward, eventually leaving the domain through the inner boundary.

We conducted test simulations with damping boundary conditions applied to all quantities at the inner radial boundary and found no qualitative differences compared to our fiducial setup, particularly in the formation of dust clumps within large-scale vortices, which is central to this study.

Simulations are carried out in cylindrical coordinates as defined in Section \ref{sec:model}. Units are such that $r_{0}=M_{*}=G=1$.
We adopt $h_{0}=H_{g0}/r_{0} =0.1$ in all runs. 
In our fiducial setup, the numerical grid spans $0.7\leq r\leq 1.3$ and $-0.4 H_{g}\leq z \leq +0.4 H_{g}$, with the azimuthal range $0\leq \varphi \leq 2\pi$ in the 3D simulations. Compared to \hyperlinkcite{lehmann2024}{Paper I}, the vertical domain has been expanded by a factor of 1.6, while the radial domain has been reduced by a factor of 0.6 to maintain manageable computational costs. The increased vertical domain size was chosen based on isothermal 2D test simulations including dust, conducted for different dust parameters ($\tau$ and $Z$). These tests confirmed that a stable, quasi-steady turbulent state is established and maintained for at least 1000 orbits. In contrast, simulations with smaller vertical domains often exhibited an artificial enhancement of turbulence after several hundred orbits, likely caused by back-reaction effects through the vertical boundaries.
The grid resolution used in our simulations is  $N_{r} \times N_{z} \times N_{\phi} = 900 \times 120 \times 628$ . This corresponds to radial and vertical resolutions of  $150/H_{0}$ , while the azimuthal resolution, at approximately  $10/H_0$ , is significantly lower. The reduced azimuthal resolution is a compromise to limit computational costs, as the structures we expect to form in our simulations are predominantly axisymmetric or elongated in the azimuthal direction. Compared to \PaperI, where the radial and vertical resolution was  $200 /H_0$ , the current resolution is slightly lower. However, in pure gas simulations, we did not observe any significant differences in the results. Additionally, this adjustment considerably reduces the memory requirements of the simulations.

The 3D simulations are conducted on a GPU cluster, which is crucial for efficiently handling the high spatial resolution. Most simulations span 1,000 reference orbits, with select cases extended further as needed. Instances of longer simulation times will be explicitly noted.

 \subsection{Diagnostics}\label{sec:diagnostics}

\subsubsection{Turbulence properties}

As in \hyperlinkcite{lehmann2024}{Paper I}, we describe the radial turbulent angular momentum transport in 3D simulations via the dimensionless quantity
\begin{equation}\label{eq:alpha}
\alpha_{r}(t) = \frac{\langle \rho_g v_{g,r} \delta v_{g,\varphi } \rangle_{r\varphi z}}{\langle P \rangle_{r \varphi z}},
\end{equation}
where the brackets denote averaging over spatial dimensions as indicated in the subscript, and $\delta v_{\varphi}$ is the azimuthal velocity deviation from its ground state value.  In principle, we could also define a measure for the vertical transport of angular momentum, similar to the $\alpha_z$ parameter introduced in \PaperI. However, this is not particularly useful in our current context since we can directly measure the dust scale height, which is a more directly relevant  quantity for describing the impact of turbulence on the dust layer, which is the focus of this study.

To quantify the turbulent vertical stirring of dust, we define the root mean squared (RMS) vertical dust velocity as:
\begin{equation}
\text{RMS}(v_z) \equiv \sqrt{\langle v_{d,z}^2 \rangle_{r \varphi z}},
\end{equation}
where the subscript $d$ explicitly denotes dust velocities. 

Throughout this work, unless stated otherwise, velocity quantities without subscripts refer to the dust component. We note that for $\tau \ll 1$, dust and gas are tightly coupled, meaning that their velocity differences are generally small.

The dust layer thickness is quantified via the dust scale height $H_d$, which is obtained by fitting a Gaussian profile as described in Eq.~(\ref{eq:dustdens}) to the radially and azimuthally averaged dust density. If necessary, a vertical shift parameter is included in the fit to account for cases where the dust layer is locally displaced from the midplane ($z=0$).

Spatial averages of quantities presented in this work are computed over a well-defined domain to ensure consistency. The radial averaging domain is chosen as $0.9 < r/r_0 < 1.2$ to mitigate boundary effects. This asymmetric domain is chosen because the inner boundary is generally subject to increased hydrodynamic activity. Azimuthal averaging is performed over the full $2\pi$ domain. Vertical averaging is performed over the entire computational domain, except for RMS$(v_z)$, which is averaged only over a few grid cells above and below the midplane ($z=0$). This choice reflects the fact that the dust layer is generally settled, making it uninformative to measure turbulent stirring far from the midplane.

%%%%%%%%%%%%%%%%%%%%%%%%%%%%
 \begin{figure*}
 \centering 
 	\includegraphics[width= 0.95 \textwidth]{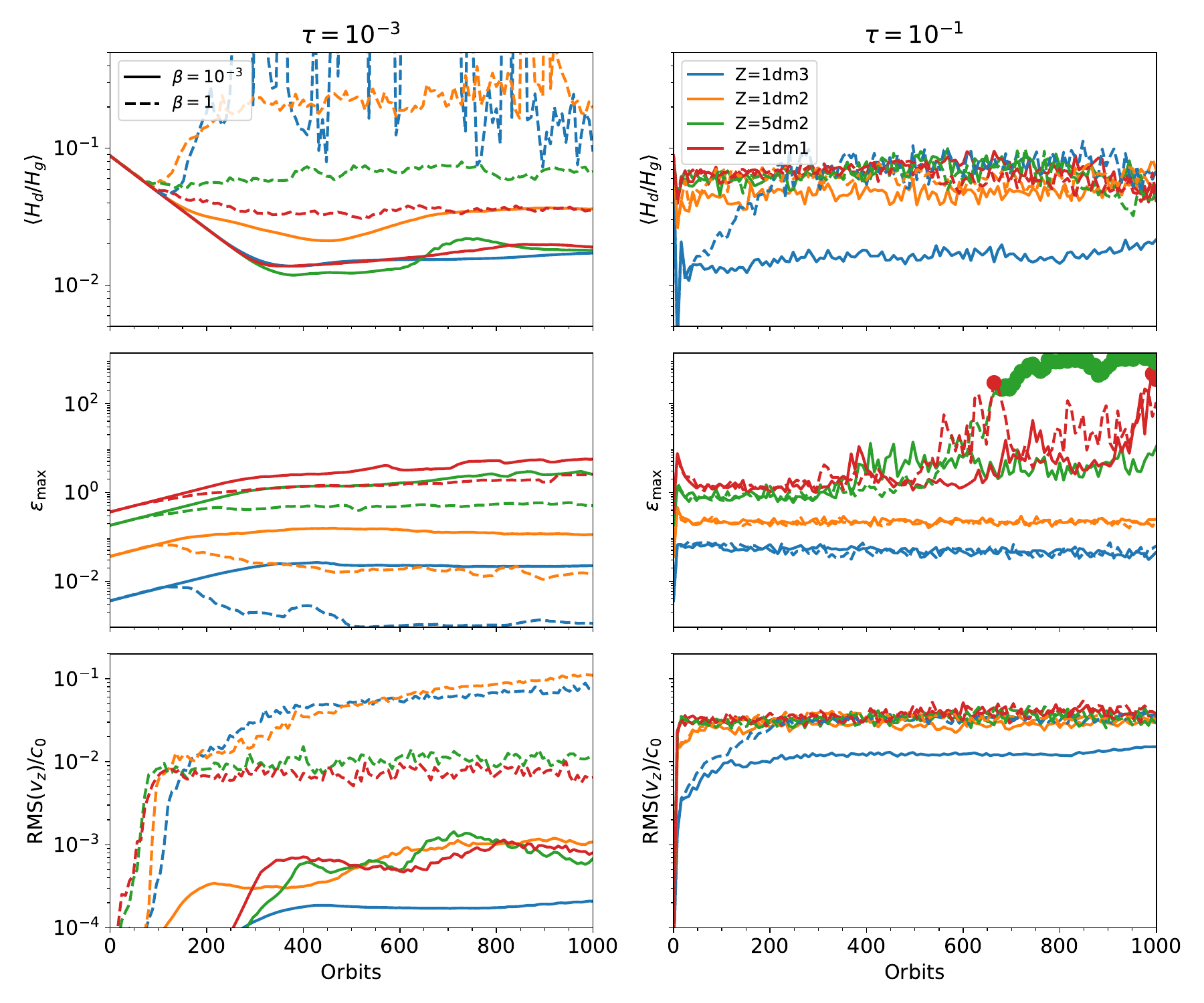}
     \caption{Dust scale height, maximum dust-to-gas density ratio, and RMS vertical dust velocities from 2D axisymmetric simulations with cooling time $\beta=0.001$ (solid curves) and $\beta=1$ (dashed curves). The simulations include metallicities ranging from $[0.001 - 0.1]$, with Stokes numbers $\tau=0.001$ (left panels) and $\tau=0.1$ (right panels), and adopt $p=1.5$ and $q=2$.}
     \label{fig:results_2d}
 \end{figure*}
%%%%%%%%%%%%%%%%%%%%%%%%%%%%

\subsubsection{Strong dust clumping}

Following \citet{li2021}, we define “strong dust clumping” in our non-selfgravitating simulations as any region where the local dust density $\rho$ exceeds the “Roche density” \citep{goldreich1973}:
\begin{equation}
\rho_{\mathrm{Roche}} = \frac{9\,\Omega_K^2}{4\pi\,G}.
\end{equation}
Surpassing $\rho_{\mathrm{Roche}}$ suggests that, in principle, dust self-gravity could overcome the stellar tidal force if it were included. Indeed, several of our simulations achieve $\rho \gg \rho_{\mathrm{Roche}}$. However, $\rho>\rho_{\mathrm{Roche}}$ alone does not guarantee true collapse: shear, gas pressure, and turbulence can still disrupt clumps.

A commonly used parameter to assess axisymmetric gravitational stability in a Keplerian flow is the hydrodynamic analogue of the Toomre parameter \citep{toomre1964}:
\begin{equation}
Q = \frac{c_s \,\Omega_K}{\pi G\,\Sigma_g},
\end{equation}
which balances shear (via $\Omega_K$) and thermal pressure (via $c_s$) against self-gravity (via $\Sigma_g$). In a pure-gas disk, $Q<1$ indicates susceptibility to axisymmetric collapse, while $Q>1$ implies stabilization. When dust is significant, a modified criterion $Q < (1 + Z)^{3/2}$ 
\citep{takahashi2014} includes both the added dust mass and changes in effective pressure. Under such an instability, radial perturbations can grow until densities naturally exceed $\rho_{\mathrm{Roche}}$.

However, large $Q$ does not forbid local dust trapping (e.g.\ in vortices or via streaming instability), which can produce small-scale concentrations that surpass $\rho_{\mathrm{Roche}}$ and become tidally bound, provided local shear and pressure are also overcome. The Toomre analysis assumes nearly uniform background profiles and small perturbations, so it does not fully capture these nonlinear processes. Demonstrating actual collapse in such traps typically requires self-gravitating simulations or a detailed analysis of the competition between clump growth and disruption timescales \citep[e.g.,][]{carrera2021a,  youdin2002, schafer2024}.

In short, since our simulations do not include self-gravity, we use $\rho > \rho_{\mathrm{Roche}}$ as a practical indicator of strong clumping, acknowledging that additional stabilizing forces can still prevent collapse. In realistic disks, even if $Q>1$, lower $Q$ values reduce shear stabilization, making it easier for overdensities to remain above $\rho_{\mathrm{Roche}}$ \citep{takahashi2014}.

\section{Results of 2D axisymmetric simulations}\label{sec:sim_2d}

To gain initial insights into the problem, we begin by presenting results from 2D simulations. These simulations simplify the analysis by excluding the complexity introduced by vortices and additional instabilities, such as the elliptic instability \citep{lesur2009} or the Rossby wave instability (RWI: \citealt{lovelace1999}).

\subsection{Dust concentration and turbulence}

Figure \ref{fig:results_2d} presents the averaged dust-to-gas scale height ratio $\langle H_{d}/H_{g} \rangle$, the maximum value of the dust-to-gas density ratio $\epsilon$, and the root-mean-square (RMS) vertical velocities from 2D axisymmetric simulations. Dashed curves correspond to simulations with a cooling time $\beta = 1$, while solid curves represent $\beta = 10^{-3}$, which effectively corresponds to the isothermal limit $\beta\to 0$. The left panels show results for small grains with $\tau = 10^{-3}$, and the right panels for large grains with $\tau = 0.1$.

A key observation is that significant turbulence levels are present in all simulations, including those with $\beta = 10^{-3}$. This is evident from the RMS$(v_z)$ values, which reach a significant fraction of the sound speed $c_{s0}$. This indicates that the COS cannot be the sole source of turbulence, as it vanishes in the isothermal limit (e.g. \citealt{lehmann2023}). We hypothesize that, in addition, the VSSI develops in these simulations. The VSSI is expected because the stratified dust layer induces vertical shear in the dust-gas mixture, which drives the instability (see Section \ref{sec:vssi} for further discussion). For instance, in simulations with $\tau = 0.1$, the initial rapid increase (or decrease) of $\epsilon$ or $\langle H_{d}/H_{g} \rangle$, driven by dust settling, stalls after several orbits. This is followed by a puffing-up of the dust layer, likely due to turbulence generated by the VSSI, leading to a quasi-steady state of the dust-gas mixture.  

 \begin{figure}
 \centering 
 	\includegraphics[width= 0.5 \textwidth]{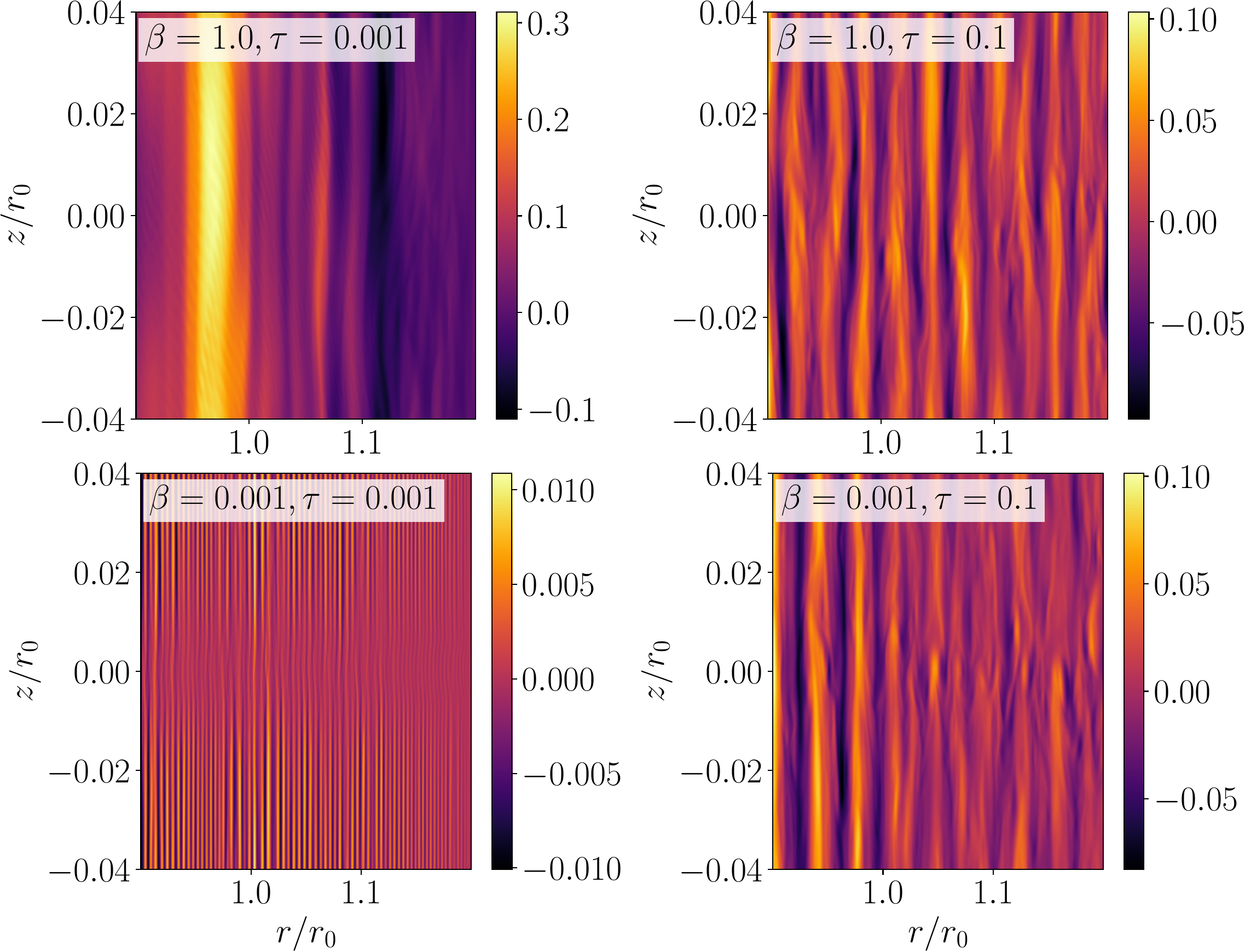}
     \caption{Snapshots of the vertical gas velocity in the saturated state of 2D axisymmetric simulations with $Z=0.01$ around 1000 orbits. The upper and lower panels compare different cooling times, while the left and right panels compare different Stokes numbers. The upper left panel shows the characteristic saturation of the COS with coherent elevator flows (cf. \PaperI). The remaining panels show the saturation of the VSSI, as explained in the text.}
     \label{fig:vssi_cos}
 \end{figure}

We revisit this in Section \ref{sec:vssi}, where we provide arguments as to why this early onset of turbulence is unlikely to be caused by the classical SI. It is worth noting, however, that the SI is still expected to occur at later stages of our simulations, when the VSSI, and possibly the COS (for $\beta = 1$), have already reached a nonlinearly saturated state. In particular, the strong clumping observed in some simulations discussed here is expected to result from the SI.

%%%%%%%%%%%%%%%%%%%%%%
 \begin{figure*}
 \centering 
 	\includegraphics[width= 0.85 \textwidth]{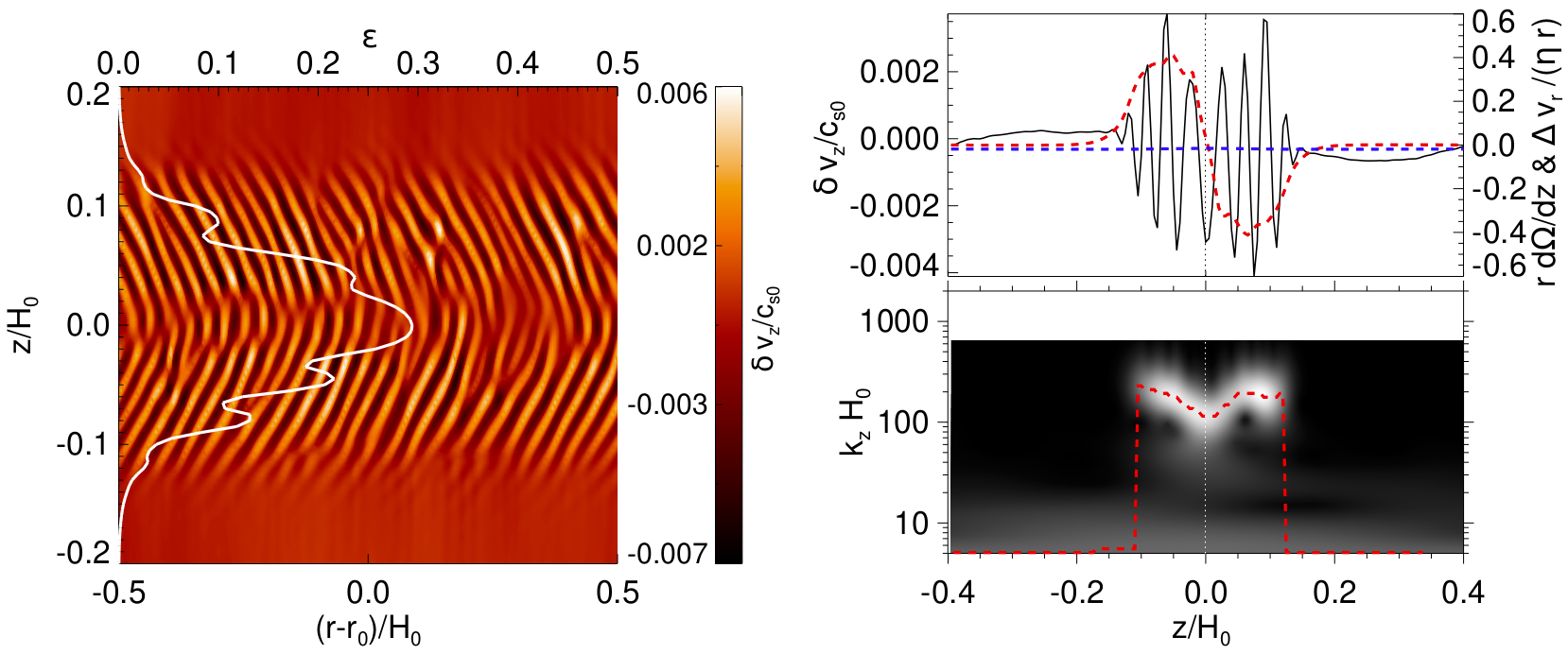}
     \caption{Illustration of the emergence of the VSSI in a 2D simulation with $\tau=0.01$, $Z=0.05$, and $\beta=0.001$. The left panel shows the vertical gas velocity around eight orbits, exhibiting a prominent wave pattern associated with the dust layer (white curve), radially averaged around $r = r_0$. The right panel shows vertical profiles of the gas velocity (black solid), vertical shear (red dashed), and radial drift (blue dashed). The bottom panel shows the wavenumber corresponding to the maximum wavelet power at each height.}
     \label{fig:vssi}
 \end{figure*}
%%%%%%%%%%%%%%%%%%%%%%

For simulations with $\tau = 10^{-3}$, significant differences are evident across all quantities between the two cooling times. This is plausible, as VSSI turbulence is expected to be weak for such a small Stokes number \citep{lin2021}, allowing the COS to dominate (for $\beta = 1$) across the metallicity $Z$ range considered here. In contrast, simulations with $\tau = 0.1$ exhibit less variation between the two cooling times, suggesting that the VSSI dominates, except at very low metallicity $Z = 0.001$.

Overall, a comparison of dashed and solid curves across all panels shows that the presence of the (axisymmetric) COS generally reduces $\epsilon$, as it stirs up the dust layer and counteracts vertical gravity. 

Hydrodynamic instabilities, including the COS and the VSSI, generally induce turbulent vertical motions in the dust layer and therefore contribute to vertical dust stirring. Their relative influence can be assessed by comparing the RMS$(v_z)$ curves for different cooling times. Specifically, if the RMS$(v_z)$ curves differ significantly between $\beta = 1$ and $\beta = 10^{-3}$, this suggests a strong contribution from COS, since the VSSI does not rely on gas cooling \citep{lin2021}. Conversely, if the curves are similar for both cooling times, the added effect of COS for $\beta = 1$ is negligible, implying that the VSSI dominates dust stirring.

This effect is most pronounced in simulations with $\tau = 10^{-3}$. In particular, for $Z = 0.001$, the COS fully unsettles the dust layer upon saturation. Only at higher metallicities $Z \gtrsim 0.05$ does significant dust clumping occur (indicated by the thickened curve), likely driven by the classical SI within the dust layer. We will return to this point in Section \ref{sec:discuss_axi}.

At sufficiently high metallicity, the COS appears to be largely suppressed, allowing the VSSI to dominate. Examining the RMS$(v_z)$ curves, we find that for $\tau = 10^{-3}$, COS suppression occurs at $Z \gtrsim 0.05$. For $\tau = 0.1$, suppression requires only $Z \gtrsim 0.01$. This behavior is illustrated in Figure \ref{fig:vssi_cos} for $Z=0.01$, where the contours depict the vertical velocity perturbations ($\delta v_z$) in the saturated state. For $\tau = 0.1$, the VSSI dominates, and the contours for $\beta = 10^{-3}$ and $\beta = 1$ are similar, indicating that the VSSI has reached saturation. In contrast, for $\tau = 10^{-3}$, the COS saturates into persistent elevator (and zonal) flows for $\beta = 1$. The radial VSSI length scales for $\tau = 10^{-3}$ are significantly smaller than for $\tau = 0.1$, consistent with the findings of \citet{lin2021}.

An intriguing difference in the results of Figure \ref{fig:results_2d} arises in the behavior of the dust scale height for the two Stokes numbers. For $\tau = 0.1$, $\langle H_{d}/H_{g} \rangle$ increases with increasing $Z$ in simulations with $\beta = 0.001$, whereas for $\tau = 10^{-3}$, the scale height decreases for $Z \geq 0.01$. For $\beta = 1$, a decreasing $\langle H_{d}/H_{g} \rangle$ for $\tau = 10^{-3}$ is expected, as COS dominates in this regime but weakens with increasing Z. This weakening becomes significant only for $Z \gtrsim 0.05$, where midplane values of $\epsilon$ approach unity, consistent with the linear analysis of \citet{lehmann2023}.
The overall complexity of the dust layer thickness arises from the interplay of the COS and the VSSI (and possibly the SI), whose resulting turbulence depends on the parameters $Z$ and $\tau$ in distinct ways. We will revisit this in more detail in Section \ref{sec:eddy}.

 \begin{figure*}
 \centering 
 	\includegraphics[width= 0.8 \textwidth]{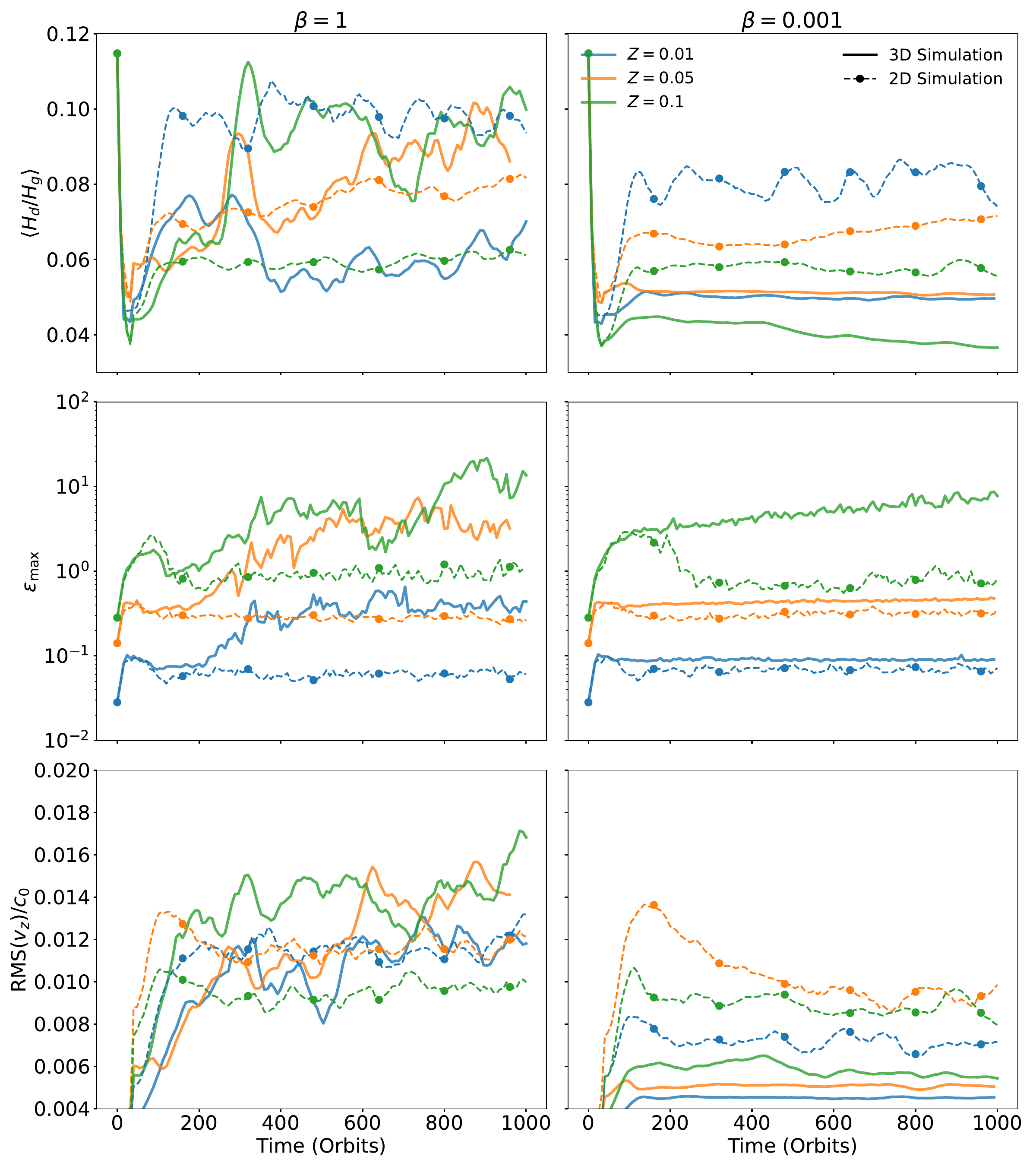}
     \caption{Comparison of diagnostic quantities (maximum dust-to-gas density ratio, dust scale height, RMS velocities)  between 2D and 3D simulations with $\tau=0.01$ and cooling times $\beta=1$ (left) and $\beta=0.001$ (right). The curves for $\langle H_{d}/H_{g} \rangle$ and $RMS(v_z)$ have been smoothed for improved visibility.}
     \label{fig:2d_3d_comp}
 \end{figure*}

\subsection{Occurrence of the VSSI}\label{sec:vssi}

To strengthen our claim that the VSSI is active in our simulations, we consider Figure \ref{fig:vssi}. The left panel shows contours of the vertical gas velocity $\delta v_z$ after eight orbital periods, revealing the onset of a linear instability associated with the dust layer, as illustrated by the white curve showing $\epsilon(z)$. The thin dust layer induces a vertical shear $r \mathrm{d}\Omega /\mathrm{d} z$, which can drive the VSSI \citep{lin2021}.

The upper-right panel shows a vertical profile of the radially averaged (across $\Delta r \sim 0.3 H_0$) vertical velocity $\delta v_{z}$ (black solid curve) at eight orbital periods. Also shown are the vertical shear $r \mathrm{d}\Omega /\mathrm{d} z$ (red dashed curve) and the radial dust-gas drift $(\delta v_{d,r} - \delta v_{g,r})/(\eta r)$ (blue dashed curve), both plotted as functions of height $z$. Radial drift, the driving force of the classical SI, is represented using $\eta r$, a characteristic length scale associated with linear SI. In our simulations, vertical shear dominates over radial drift in the perturbed region, making it unlikely that the observed instability is the classical SI.

 \begin{figure*}
 \centering 
 	\includegraphics[width= 0.82 \textwidth]{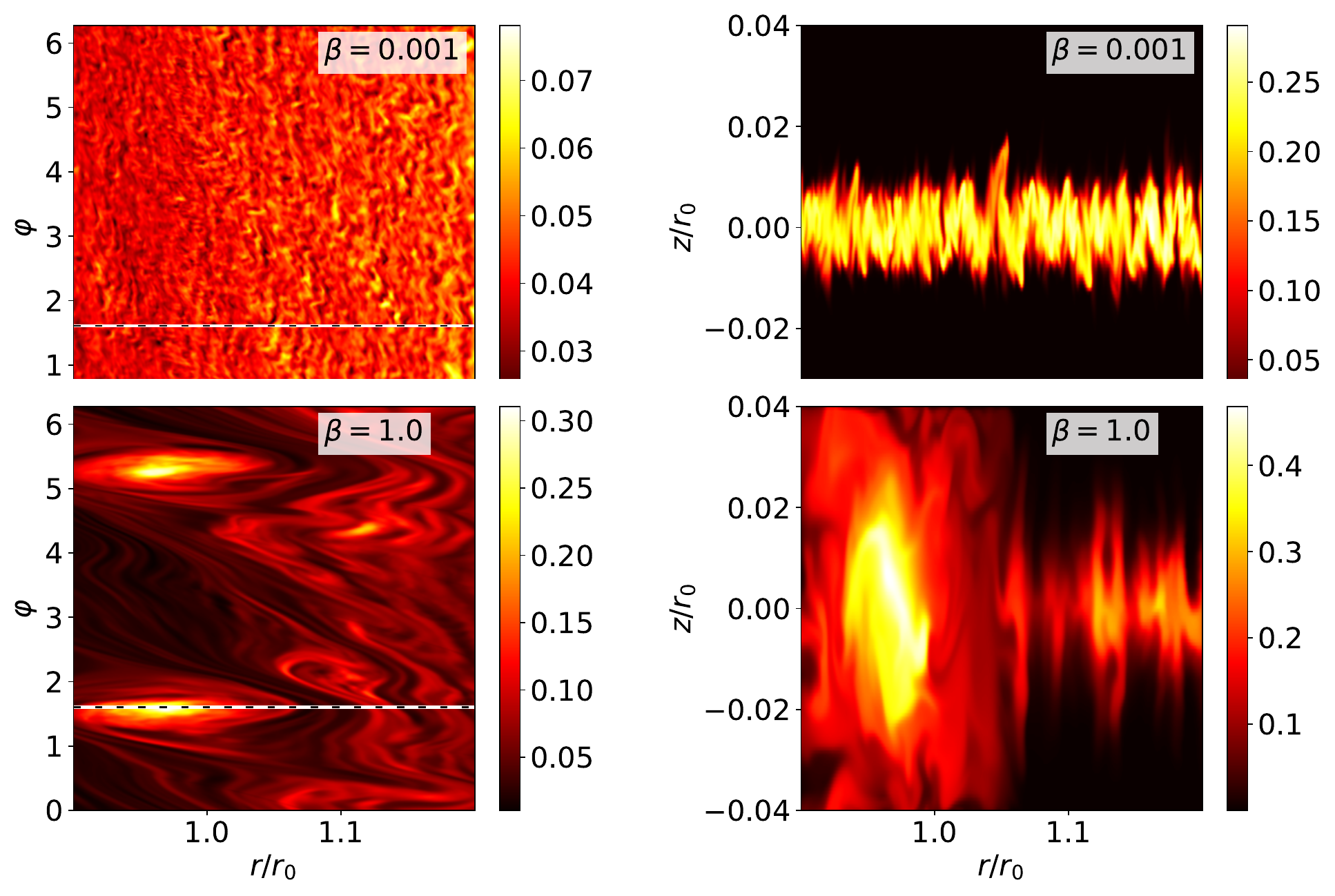}
     \caption{Contours of the dust-to-gas density ratio $\epsilon$ in the nonlinear state of 3D simulations at $t\sim 200 \, \text{ORB}$ with $Z=0.01$, $\tau=0.01$, and with cooling time $\beta=0.001$ (top) and $\beta=1$ (bottom). Left panels correspond to the disk mid-plane. Right panels correspond to the azimuth as indicated by the dashed lines in the left panels.}
     \label{fig:vssi_cos_3D}
 \end{figure*}

The bottom-right panel shows a wavelet power spectrum of $\delta v_z$ as a function of vertical wavenumber. The spectrum reveals the dominant vertical wavenumber associated with the linear perturbation. Using Eq. (58) of \citet{lin2021}, we can estimate the minimum total wavenumber required for the VSSI under our simulation conditions. Assuming $\epsilon = 0.3$ (see Figure \ref{fig:vssi}), $St = 0.01$ (here $\tau$), $H_d/H_g = 0.055$ (see Figure \ref{fig:vssi}), and $z = H_d$, we find a minimum total wavenumber $k H_0 \gtrsim 15$, which is fulfilled by a large margin, since we find $k_z H_0 \sim 100$ with $k^2=k_r^2 + k_z^2$.

\section{Results of 3D simulations}\label{sec:sim_3d}

In this section we present the results of our 3D simulations. As for our 2D simulations we restrict to cooling times $\beta=1$ (optimal regime of the COS) and $\beta=0.001$ (isothermal regime). Furthermore, we use $p=2.5$ and $q=2$ in all simulations.

\subsection{comparison between 2D and 3D simulations}

We begin by comparing the results of 2D and 3D simulations. Figure \ref{fig:2d_3d_comp} shows the maximum dust-to-gas density ratio $\epsilon_{\text{max}}$, the dust-to-gas scale height ratio $\langle H_{d}/H_{g} \rangle$, and the RMS vertical velocity $\text{RMS}(v_z)$ for 2D (dashed curves) and 3D (solid curves) simulations. The left panels correspond to $\beta = 1$, while the right panels correspond to $\beta = 0.001$, both for a Stokes number of $\tau = 0.01$.

First, we note that the turbulence strength $\text{RMS}(v_z)$ increases with increasing metallicity $Z$ in all 3D simulations. This suggests that, for the adopted $\tau = 0.01$, the VSSI contributes significantly to the turbulence, as COS turbulence alone is expected to weaken with increasing $Z$ \citep{lehmann2023}. Furthermore, we infer that in the 2D simulations, the VSSI dominates\footnote{Actually, the VSSI starts dominating already for $Z\geq 0.03$, which is not explicitly shown here} for $Z \gtrsim 0.05$, as indicated by the similar $\text{RMS}(v_z)$ values for the two cooling times. In the 3D simulations, however, the COS appears more vigorous, substantially enhancing the resulting turbulence.

We also observe substantially higher dust-to-gas ratios $\epsilon_{\text{max}}$ in the 3D simulations, primarily due to the formation of vortices (see Section \ref{sec:vortices}). Unlike in 2D, these large-scale, coherent vortices can trap dust both radially and azimuthally, further enhancing dust concentration—with amplification factors reaching $\sim 60\text{–}70$. This is illustrated in Figure \ref{fig:vssi_cos}, which shows planar and meridional contours of $\epsilon$ for 3D simulations with $Z = 0.01$ and different cooling times. The VSSI produces weak, small-scale non-axisymmetric structures in these simulations, corresponding to low vorticity production. In contrast, simulations with $\beta = 1$ are dominated by the COS, leading to the formation of large-scale, long-lived vortices that migrate radially inward.

%%%%%%%%%%%%%%%%%%%
\setlength{\textfloatsep}{10pt plus 1.0pt minus 2.0pt}
\begin{figure*}[ht] % 't' indicates top placement
\centering 
    \includegraphics[width=\textwidth]{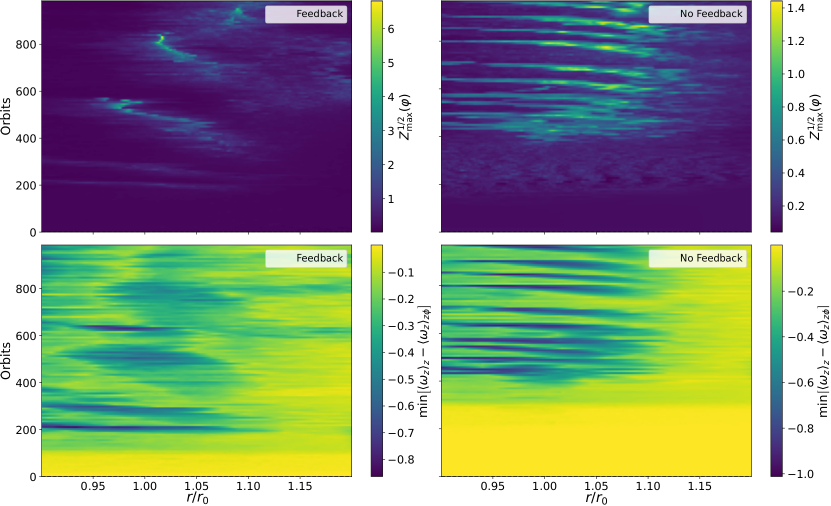}
    \caption{Effect of dust feedback on dust concentration and vorticity in vortices for $\beta = 1$, $Z = 0.01$, and $\tau = 0.05$. The left panels show results with dust feedback, while the right panels show results without feedback. The upper panels display the maximum azimuthally averaged dust-to-gas ratio $Z_{\max}^{1/2}(\varphi)$ as a function of radius and time, illustrating how dust accumulates within vortices. The lower panels show the minimum vertical vorticity deviation $\min[(\omega_z)_{z} - (\omega_z)_{z\varphi}]$, highlighting the structure and evolution of large-scale vortices. With feedback, vortices accumulate significant amounts of dust, leading to stronger dust clumping and reduced migration rates. In contrast, without feedback, multiple vortices persist, but dust concentration remains lower, and migration is more rapid.}
    \label{fig:3d_feedback_1}
\end{figure*}
%%%%%%%%%%%%%%%%%%%

In the isothermal simulations with $Z = 0.1$, as shown in Figure \ref{fig:2d_3d_comp}, a dust ring forms early on, as seen by the strong rise in $\epsilon_{\max}$ in the upper right panel. In the 2D case, the ring dissolves, whereas in the 3D case, it undergoes weak clumping, likely driven by the SI. Yet, even in these simulations, strong clumping ($\rho \geq \rho_{\text{Roche}}$) does not occur within 1000 orbits.

The behavior of the dust layer thickness $H_d$ remains complex, with no clear overall trend, likely due to the interplay with the VSSI. Notably, in simulations with $\beta = 1$, the behavior of $H_d$ differs between 2D and 3D simulations, showing opposite trends. Additionally, both $H_d$ and $\text{RMS}(v_z)$ in the 2D simulations differ from the results in Section \ref{sec:sim_2d}, where different Stokes numbers were used.

Interestingly, the vertical turbulence driven by the VSSI appears to be more vigorous in 2D simulations than in 3D simulations, as seen in the lower right panel. However, $RMS(v_{\varphi})$ (not shown here) is substantially larger in 3D simulations. This is likely due to the presence of small-scale, non-axisymmetric structures present in 3D simulations. These differences underscore the distinct dynamics of turbulence and dust-gas interactions between 2D and 3D simulations, emphasizing the importance of dimensionality in capturing the full complexity of these processes.

%  Given that the VSSI is the origin of the turbulence, we assume that the correlation time of turbulent fluctuations \citep{yl07} $t_{eddy}\sim 1/(\partial v_{\varphi} / \partial z)\ll 1/\Omega$. On the other hand, we have $t_s \Omega \ll 1$. Thus, based on Figure 1 in \citet{yl07} we are in either regime VI or VII, for which these authors provide the expressions $H_d/H_g = \sqrt{\alpha_z/\tau^2 \, t_{eddy} \Omega}$ and $H_d/H_g = \sqrt{\alpha_z /\tau}$. Since the vertical equilibrium shear increases with increasing $Z$ \citep{lin2021}, we expect $t_{eddy}$ to  \emph{decrease} accordingly.
% The additional difficulty in the present situation is that $\alpha_z$ is not constant, but will also change with $Z$, and a precise understanding of the behavior of the dust layer thickness is beyond the scope of this paper. 

 %  \begin{figure}
 % \centering 
 % 	\includegraphics[width= 0.5 \textwidth]{N2_dust_density_profile_beta0.001.pdf}
 %     \caption{}
 %     \label{fig:Nz2}
 % \end{figure}

\subsection{Dust concentration in vortices}\label{sec:vortices}

\subsubsection{Role of dust feedback}\label{sec:feedback}

We find that dust feedback is essential for achieving high dust-to-gas density ratios in vortices. This is illustrated in Figure \ref{fig:3d_feedback_1} for simulations with $\tau = 0.05$ and $Z = 0.01$ (and $\beta=1$), which shows contours of the metallicity $Z$ and the quantity $\text{min}[\langle \omega_{z} \rangle_{z} - \langle \omega_z \rangle_{z\varphi}]_{\varphi}$. The latter highlights the presence of large-scale vortices (see \PaperI), where $\omega_z$ is the vertical component of vorticity, defined as
\begin{equation}\label{eq:vorticity}
\boldsymbol{\omega} = \nabla \times \boldsymbol{v}. 
\end{equation}
Substantially larger values of $Z$ (approximately ten times higher) and, consequently, $\epsilon$, are achieved in the simulation with dust feedback. In this case, vortices are less numerous, their migration slows down, and they weaken as they accumulate more dust. These results suggest that dust tends to suppress vortex formation via the COS. The slowdown in migration is likely tied to the weakening of the vortices, as weaker vortices excite less intense spiral density waves and thus induce weaker radial angular momentum transport (see \PaperI). In principle, vortex weakening should be accompanied by an increase in their aspect ratio \citep{paardekooper2010}. 

However, measuring the precise aspect ratio is challenging due to the presence of multiple interacting vortices, which distort the vortex streamlines. In some cases where vortices are well-separated and coherent, more precise measurements are possible (as seen below). Apart from these special cases, we provide rough estimates to assess whether the aspect ratio remains below the critical threshold for the elliptic instability ($\chi < 4$).

We do not find clear evidence for the complete disruption of vortices due to dust loading, i.e., disruption across the entire gas column. While Figure \ref{fig:3d_feedback_1} may give the impression that vortices disappear once they accumulate sufficient dust, our results indicate that dust clumps are expelled from heavily dust-laden vortices before the vortices are fully dissolved.

An example is the vortex in Figure \ref{fig:3d_feedback_1} around $r/r_0=1$ at approximately 800 orbits (left panel). Despite being heavily loaded with dust, the vortex remains intact for several tens of orbits. By 824 orbits, the dust clump has been expelled, yet the vortex column itself remains visible, indicating its continued presence.

\begin{figure}
 \centering 
 	\includegraphics[width=  0.5\textwidth]{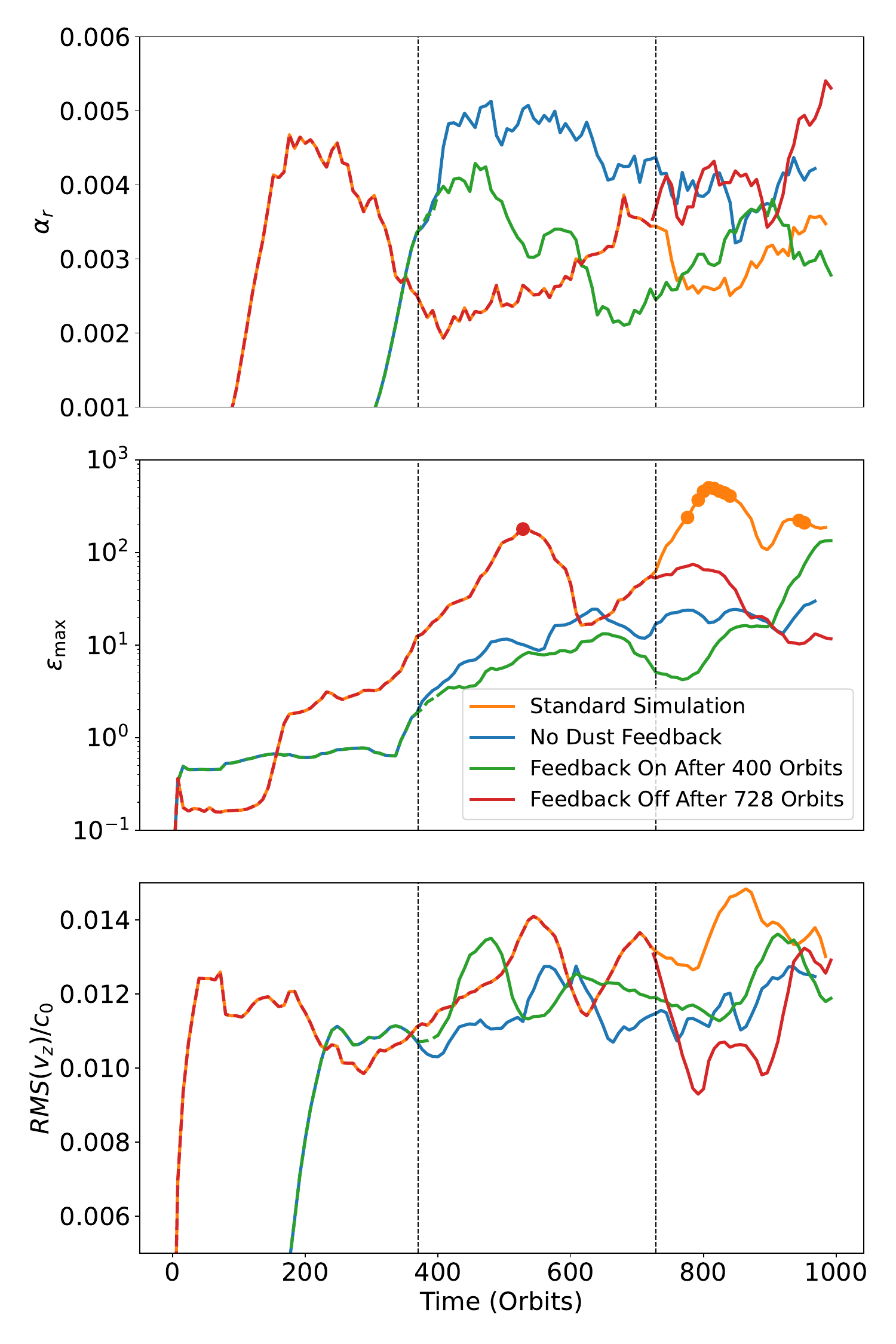}
     \caption{Effect of feedback on $\alpha_r$, $\epsilon_{\text{max}}$, and RMS($v_z$) in simulations from Figure 6. Additional curves show feedback switched off (on) at 728 (400) orbits. Feedback alters the vortex aspect ratio $\chi$, turbulence, and clumping, as discussed in the text. Curves are smoothed for clarity.}
     \label{fig:3d_feedback_2}
 \end{figure}

Figure \ref{fig:3d_feedback_2} presents the curves for $\alpha_r$, $\epsilon_{\text{max}}$, and $\text{RMS}(v_z)$ for the same set of simulations, including cases where feedback was “switched” on after 400 orbits and off after 728 orbits. While the impact of feedback is not dramatic, it is clearly observable in all curves. Specifically, turning off feedback causes an immediate increase in $\alpha_r$, while $\epsilon_{\text{max}}$ and $\text{RMS}(v_z)$ both decrease. The moderate decline in $\text{RMS}(v_z)$ is likely due to the suppression of the VSSI. Conversely, when feedback is switched on, the trends reverse, though the changes are slightly less pronounced.

We find (not shown) that the aspect ratio of the dominant vortex slightly decreases when feedback is turned off (red curve), from $\chi \sim 6.5$ to $\chi \sim 5$ within 16 orbits. Conversely, in the simulation with feedback (orange curve), the aspect ratio increases to $\chi \sim 8$ within 16 orbits. Additionally, switching off feedback results in a decrease in turbulent activity within the vortex, suggesting that drag instabilities are responsible for sustaining the turbulence (see section \ref{sec:instabilities}).

% \setlength{\textfloatsep}{10pt plus 1.0pt minus 2.0pt}
% \begin{figure*}[ht] % 't' indicates top placement
% \centering 
%     \includegraphics[width=0.9\textwidth]{vorticity_epsilon_sequence.pdf}
%     \caption{Illustration of a vortex in the same simulation as shown in Figure \ref{fig:3d_feedback_1}, undergoing strong dust clumping with $\rho > \rho_{\text{Roche}}$. The left panels display the vertically averaged vertical component of vorticity, $\langle \omega_z \rangle_z$, while the right panels show $\log(\epsilon)$ at the disk midplane. Vortices are distorted by interactions with neighboring vortices. The aspect ratio of the dominant vortex yields $\chi \sim 8-10$, making it unlikely that the fine structures visible within the vortex are caused by elliptic instability. Instead, they are more likely the result of drag instabilities. The dense dust clump is eventually expelled from the vortex, as shown in the final snapshot, after which the vortex rapidly dissolves.}
%     \label{fig:3d_feedback_1b}
% \end{figure*}

\subsubsection{Strong clumping and vortex suppression}

 \begin{figure}
 \centering 
 	\includegraphics[width=  0.5\textwidth]{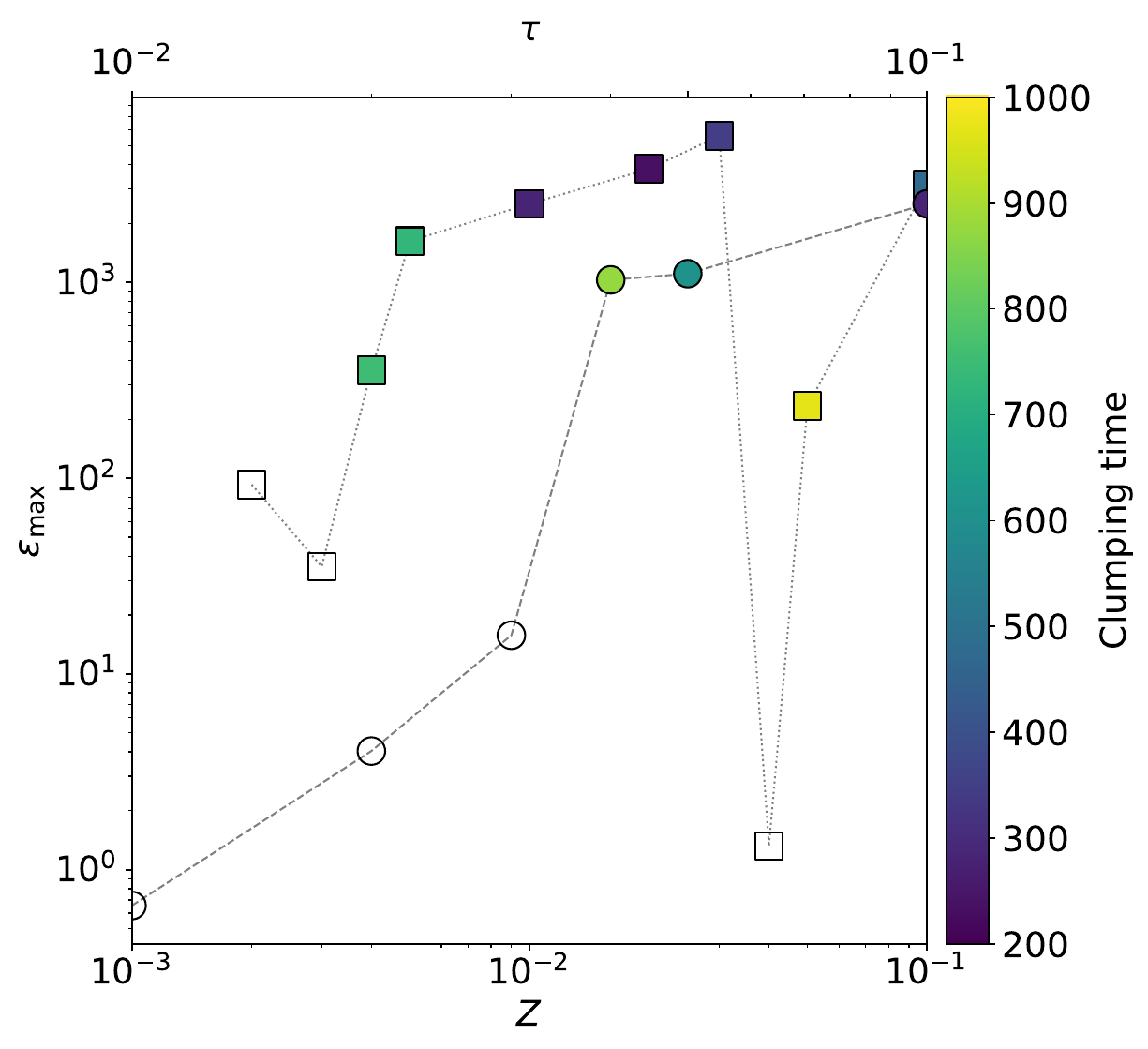}
     \caption{Achieved maximum values of the dust-to-gas ratio in two series of simulations. The circles denote simulations with $Z=0.01$ and increasing $\tau$, whereas the squares represent simulations with $\tau=0.1$ and increasing $Z$. Filled symbols indicate simulations exhibiting strong clumping. The clumping time (time until $\rho>\rho_{Roche}$ is achieved first during the simulation) is indicated by the colors (in orbits).}     \label{fig:cos_clumping}
 \end{figure}

 \begin{figure}
 \centering 
 	\includegraphics[width=  0.5\textwidth]{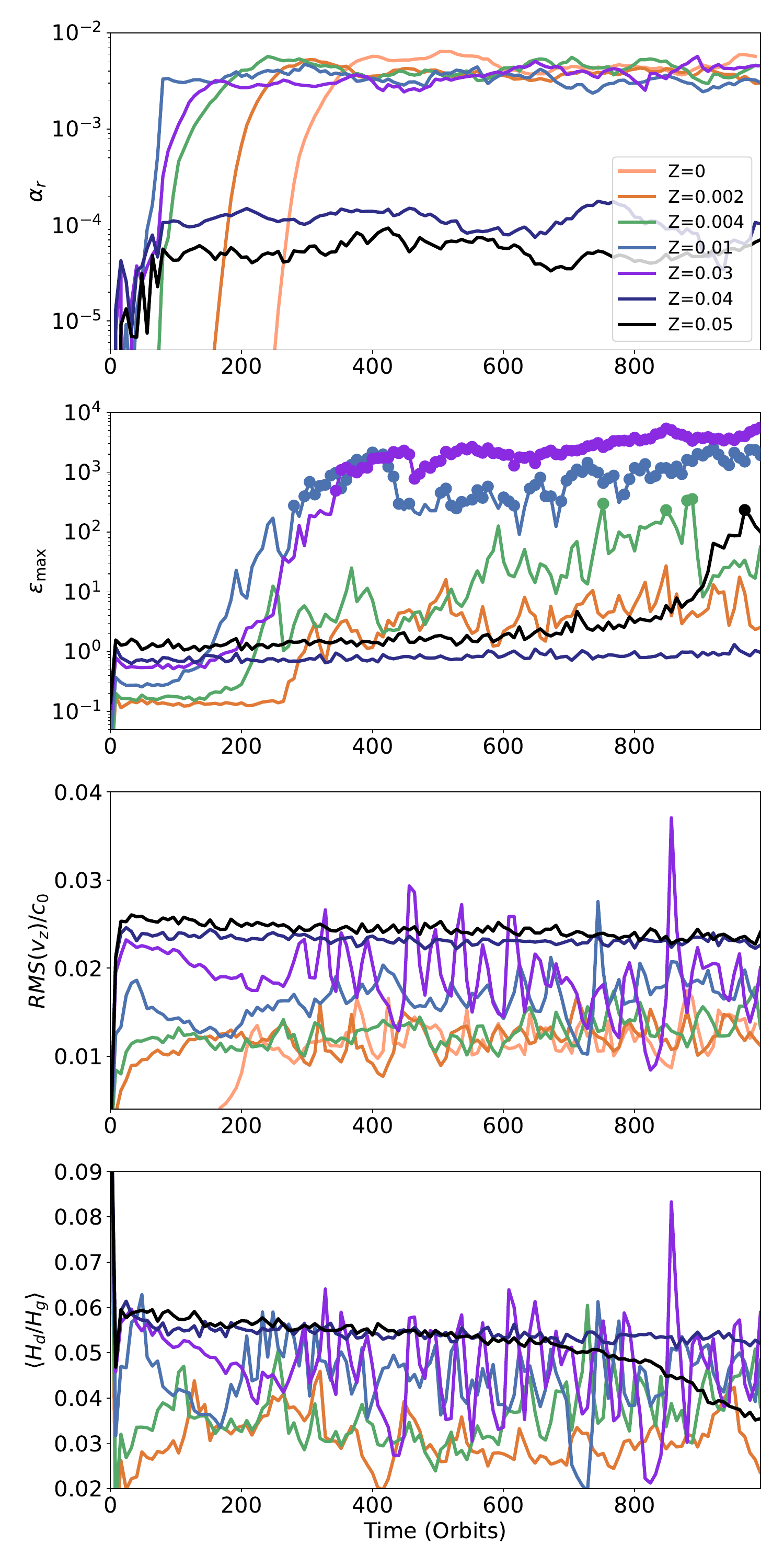}
     \caption{Measured values of $\alpha_r$, $\epsilon_{\text{max}}$, $\text{RMS}(v_z)$, and $\langle H_{d}/H_{g} \rangle$, for a series of 3D simulations with $\tau = 0.1$ and increasing $Z$. A noticeable drop in $\alpha_r$ is observed at $Z = 0.04$, indicating the suppression of large-scale vortices and the associated angular momentum transport. Consequently, strong clumping such that $\rho>\rho_{\text{Roche}}$ (indicated by the filled circles), as observed for $0.004 \leq Z \leq 0.03$, is absent within 1000 orbits. For $Z \geq 0.05$, clumping occurs around 1000 orbits via classic SI in an axisymmetric dust ring.}
     \label{fig:cos_mitigation_1}
 \end{figure}

In the previous section, we observed simulations exhibiting strong dust clumping within COS vortices, with $\rho > \rho_{\text{Roche}}$. We also found that dust feedback can suppress vortex formation by the COS. In this section, we aim to determine the critical Stokes number for strong clumping in COS vortices at initially solar metallicity ($Z = 0.01$). Additionally, we seek to identify the critical metallicity for both strong clumping and vortex suppression at a Stokes number of $\tau = 0.1$. Note that for $\tau = 0.01$, large-scale vortices still form even at $Z = 0.1$. This is indicated by the large values of $\epsilon_{max}$ in 3D simulations compared to 2D simulations shown in Figure \ref{fig:2d_3d_comp} (upper left panel).

The results of our simulation survey are summarized in Figure \ref{fig:cos_clumping}. Squares represent simulations with varying $Z \geq 0.002$ (plotted on the lower axis) and fixed $\tau = 0.1$, while circles represent simulations with varying $\tau \geq 0.01$ (plotted on the upper axis) and fixed $Z = 0.01$. Both series of simulations assume $\beta=1$. The values of $\epsilon$ have been averaged over the final 500 orbits in each simulation. Colored symbols indicate simulations that underwent strong clumping, while uncolored symbols indicate those that did not. The color of the symbols reflects the clumping time, i.e., the time at which $\rho > \rho_{\text{Roche}}$ first occurred. Note that the square corresponding to $Z = 0.01$ represents the same simulation as the circle corresponding to $\tau = 0.1$.

 \begin{figure*}
 \centering 
 	\includegraphics[width=  \textwidth]{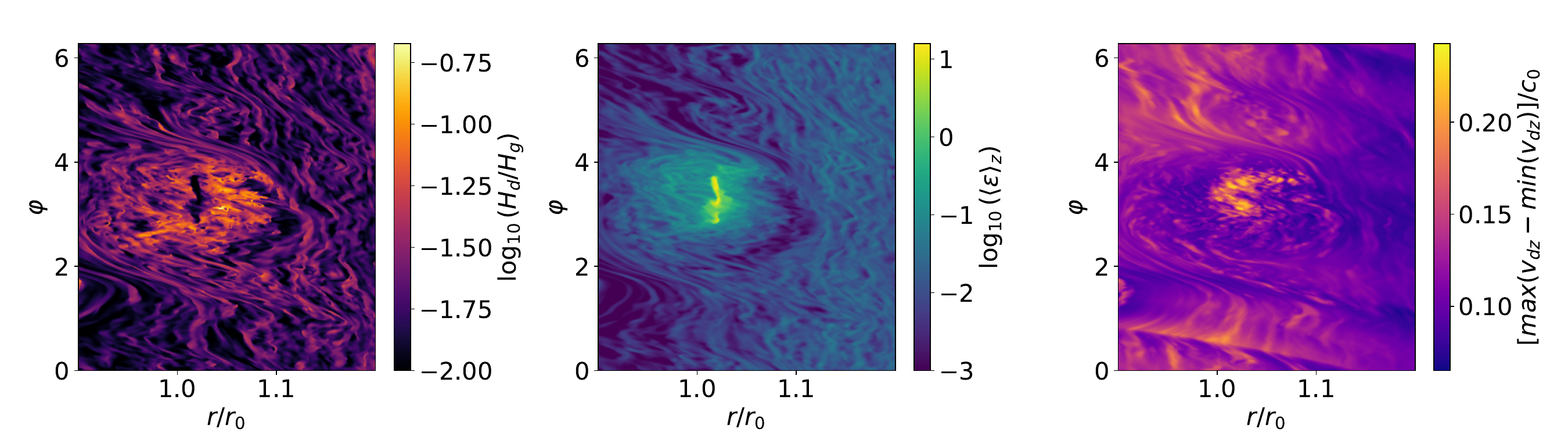}
     \caption{An example of a dust-laden vortex in the 3D simulation with $\tau = 0.1$ and $Z = 0.01$, exhibiting internal turbulence likely driven by drag instabilities. This turbulence leads to an elevated dust layer thickness within the vortex compared to its surroundings, with only the dense clump at the vortex center remaining fully settled. The vortex has an aspect ratio of $\chi \sim 8$. The left panel shows the normalized dust scale height, illustrating the thickening of the dust layer inside the vortex. The middle panel presents the dust-to-gas ratio, highlighting strong dust concentration in the vortex center. The right panel displays the normalized range of vertical dust velocity fluctuations, revealing turbulent motions within the vortex. }     \label{fig:vortex_scaleheight}
 \end{figure*}
 
At fixed $\tau$, we find that clumping occurs earliest at near-solar metallicity ($Z \sim 0.01\text{–}0.03$), where vortices are still prominent. For $Z \geq 0.04$, vortex formation is largely suppressed (we will revisit this point in section \ref{sec:disc_nonaxi}). Consequently, no clumping occurs for $Z = 0.04$, while at $Z \geq 0.05$, clumping occurs late and within an axisymmetric dust ring, likely driven by the classic SI. However, at very large metallicity (Z=0.1), clumping occurs again on a comparable timescale as at near-solar metallicity. Furthermore, we identify a critical Stokes number of approximately $\tau_{\text{crit}} \sim 0.04$ for strong clumping at initial solar metallicity ($Z = 0.01$), as indicated by the transition from open to filled circles in Figure \ref{fig:cos_mitigation_1}.

The plot reveals two critical metallicity values for $\tau = 0.1$. The critical metallicity for strong clumping is $Z = 0.004$, while the critical metallicity for suppressing large-scale vortex formation by the COS is $Z = 0.04$.

Figure \ref{fig:cos_mitigation_1} shows curves of $\alpha_r$, $\epsilon_{max}$, $\text{RMS}(v_z)$ and $\langle H_d/H_g \rangle$ for some of the simulations with $\tau = 0.1$. Here, too, we observe a clear transition when $Z$ exceeds 0.04. Specifically, $\alpha_r$ experiences a substantial drop compared to the curves at lower $Z$, indicating that radial angular momentum transport is significantly reduced in the absence of large-scale vortices, as expected. Also dust concentration weakens. On the other hand, the average vertical RMS velocities mildly increase with increasing $Z$, owing to an overall enhancement of the VSSI and possibly the SI (see section \ref{sec:instabilities}). This is also reflected in the behaviour of the dust layer thickness.

Although small, short-lived vortices continue to form for $Z \geq 0.04$, we find no evidence of strong spiral density wave excitation, which is necessary for substantial angular momentum transport. The pronounced fluctuations in $\text{RMS}(v_z)$ observed in the presence of large-scale vortices at lower $Z$ are likely related to the intermittent onset of either the elliptic instability (\PaperI), the enhancement of drag instabilities within these vortices, or a combination of both. However, in many of our results, we find that the dominant vortices have aspect ratios that are too large for a vigorous elliptic instability to develop efficiently.

This latter effect is illustrated in Figure \ref{fig:vortex_scaleheight}, which shows fine structures within the dominant vortex in the simulation with $\tau = 0.1$ and $Z = 0.01$ at around 300 orbits. These structures are associated with enhanced $\epsilon$ and result in an overall larger dust scale height within the vortex. In this case, the most likely explanation is that drag instabilities are amplified within the vortex, given the measured vortex aspect ratio of $\chi \sim 10$. Despite this, the dense dust clump at the vortex center appears to remain settled.

Additionally, it is noteworthy that dusty simulations exhibit an earlier onset of non-axisymmetric activity compared to dust-free cases. We speculate that this is due to the small but non-negligible vorticity production associated with VSSI turbulence. This vorticity increases with both increasing Z and $\tau$ (not shown) and may promote earlier large-scale vortex formation via the COS. The suppression of vortices for $Z\geq 0.04$ will be discussed in Section \ref{sec:disc_nonaxi}.

%python3 plot_dust_scale_height_new.py cos_b1d0_us_St1dm1_Z1dm2_r6H_z08H_fim053_ss203_3D_2PI_stnew_LR150 39 --rmin=0.9 --rmax=1.2  --azimuth 4.8 --right_panel 'epsilon'

\subsection{Morphology of dust structures}\label{sec:duststructures}

Here we briefly examine the degree of axisymmetry in dust structures that emerge under the influence of the COS. This investigation is particularly relevant given that one of the most striking features observed in PPDs by ALMA in recent years are rings and gaps, evident in both dust millimeter continuum and gas spectral line emissions (e.g., \citealt{vdmarel2013,vdmarel2021,alma2015,andrews2020}). Notably, only a small fraction of well studied disks display pronounced non-axisymmetric features, typically in the form of arc-shaped or crescent-shaped asymmetries, the origins of which remain a topic of active research.
As demonstrated in \hyperlinkcite{lehmann2024}{Paper I}, the COS generally leads to the formation of pressure bumps. This holds true for the simulations presented here as well, where we find, consistent with \citet{lehmann2022}, that these pressure bumps effectively trap dust and serve as favorable sites for the development of large-scale vortices.

\begin{figure*}
 \centering 
 	\includegraphics[width=  \textwidth]{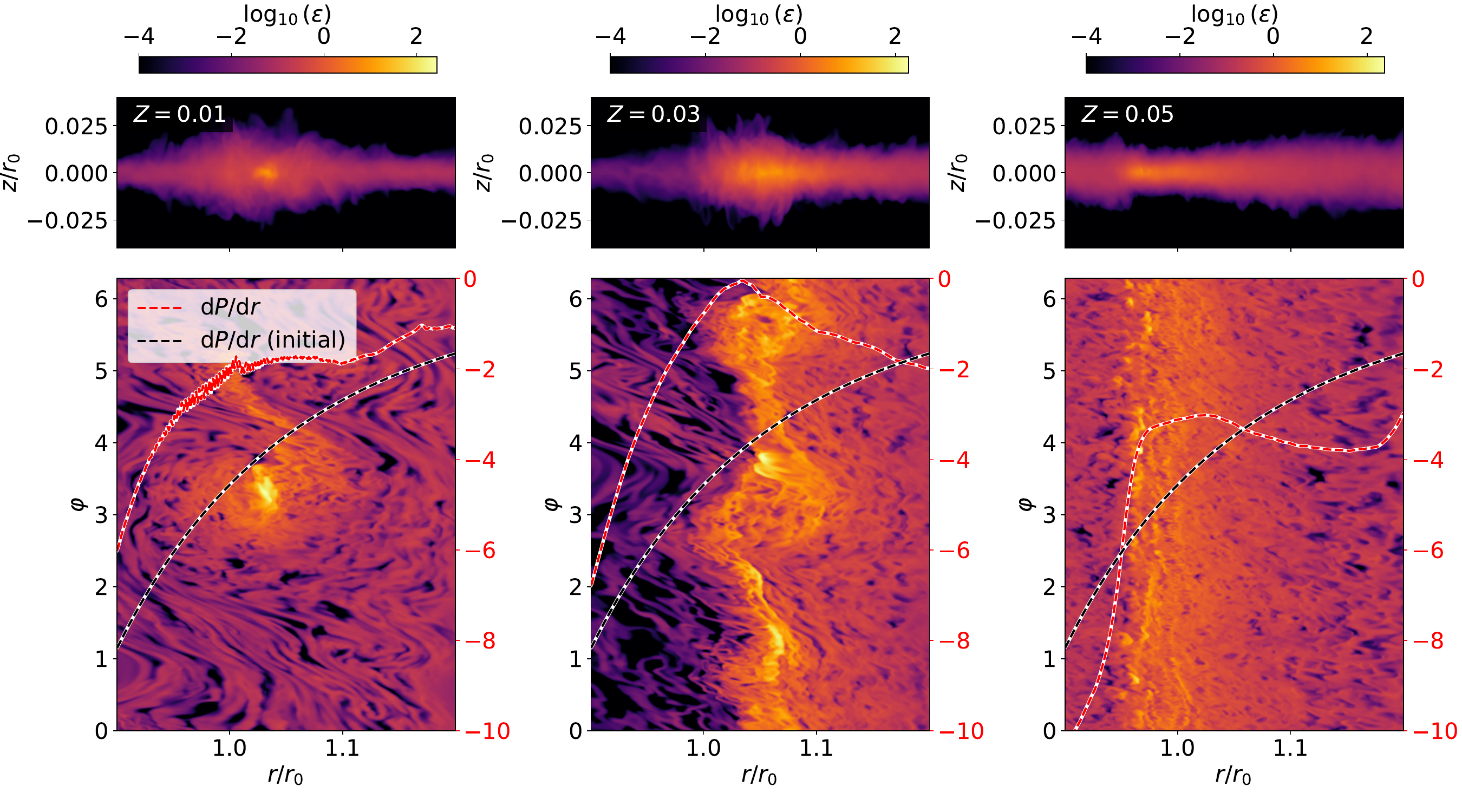}
     \caption{Contours of $\epsilon$ close to the midplane $z=0$, as explained in the text, in simulations with $\tau = 0.1$ and increasing $Z$. As $Z$ increases, the structures become progressively more axisymmetric, reflecting the enhanced suppression of large-scale vortices. The snapshots correspond to the times when the Roche density is exceeded for the first time during each simulation. The over-plotted curves show the radial pressure gradient, vertically and azimuthally averaged, at these times and at the start of the simulations.}
     \label{fig:cos_mitigation_2}
 \end{figure*}

Figure \ref{fig:cos_mitigation_2} illustrates the dust concentration, $\epsilon(z_m)$, for simulations with increasing initial metallicity $Z$, captured at the first snapshots where strong clumping occurs, marked by densities exceeding  $\rho_{\text{Roche}}$. The height $z_m$ corresponds to the maximum value of $\epsilon$ for the corresponding snapshot, and is in all cases very close to the midplane. %\mkl{to ask/clarify}
This suggests that planetesimal formation would be likely to occur in all three simulations shown, if self-gravity were included. The dashed curves in the lower panels depict the azimuthally and vertically averaged radial pressure gradient for each corresponding snapshot, as well as the initial state. Note that none of these large-scale pressure bumps are prone to the RWI, which would require much sharper bumps.

Consistent with \citet{lehmann2022}, we observe that higher background metallicity,  $Z$ , leads to a shift from forming dusty vortices to forming dusty rings — indicating increased axisymmetry in dust structures. This agreement occurs even though our simulations, in contrast to theirs, resolve drag instabilities and dense clumping. The mechanism driving this transition will be discussed in Section \ref{sec:disc_nonaxi}.

Finally, it is interesting to note that the dust-richer regions in the simulations with $Z=0.01$ and $Z=0.03$ are more puffed up than the dust-poor regions, in contrary to the simulation with $Z=0.05$.  We speculate that this difference is linked to the amplification of drag instabilities within vortices. For reference, the vortex in the $Z = 0.01$ simulation corresponds to the one shown in Figure \ref{fig:vortex_scaleheight}, captured 32 orbits earlier.

\section{Discussion}\label{sec:discussion}

\subsection{Axisymmetric COS}\label{sec:discuss_axi}

The results of our axisymmetric simulations suggest that flow structures directly resulting from the COS are not effective in significantly concentrating dust. This contrasts with the findings of \citet{lin2025}, who reported moderate enhancements of $\epsilon$ up to values of 20–30 in COS-induced zonal flows in their incompressible, viscous, unstratified shearing box simulations. The discrepancy can likely be attributed to the presence of a background pressure gradient in our simulations, quantified as $\Pi \equiv \left(\eta/h\right)_{r_0} = 0.175$. Notably, \citet{lin2025} found that even relatively modest values of $\Pi \gtrsim 0.02$ significantly hinder dust concentration in zonal flows. %\mkl{critical Pi probably also depends on cooling time, stokes number, etc. Note that the critical value of Pi=0.02 was obtained for non-optimal Peclet number. critical Pi probably increases if cooling time is optimal.}

While a direct comparison between our simulations and those of \citet{lin2025} is challenging due to differences in cooling times, vertical stratification, compressibility, and other factors, we hypothesize that the radial pressure gradient in our simulations likely plays a substantial role in suppressing dust concentration in zonal flows.

As shown by \citet{teed2021}, the negative turbulent angular momentum flux carried by the gas is responsible for zonal flow formation in their simulations. Furthermore, \citet{lin2025} observed that dust can suppress the formation of zonal flows if the angular momentum flux it carries is sufficiently positive to exceed the negative angular momentum flux associated with COS turbulence. They demonstrated that this occurs when $\epsilon \gtrsim -N_r^2/(4 \Omega_{K0}^2 \tau)$. Using Equation (\ref{eq:nr2}) and the finding in \hyperlinkcite{lehmann2024}{Paper I} that growth rates of radially global modes are comparable to local ones, we find:

\begin{equation}
\epsilon \gtrsim \frac{h_0^2 (p+q)(q+[1-\gamma]p)}{4 \gamma \tau} \left(\frac{r}{r_0}\right)^{2-q} \sim 0.1,
\end{equation}

for the parameters used in this paper. However, we observe (not shown) the formation of persistent zonal flows in all our 2D simulations, including the isothermal ones. This implies that the SI and/or VSSI also contribute to zonal flow formation.

A possible explanation for the discrepancies with \citeauthor{lin2025}’s predictions could be the presence of vertical stellar gravity in our simulations, which acts exclusively on the dust and forces it to settle into a relatively thin midplane layer. In fact, \citet{onishi2017} found that vertical self-gravity prevents dust from disrupting pressure bumps, except near the midplane, where the dust is concentrated.

In the simulation with $\beta = 1$, $\tau = 0.1$, and $Z = 0.05$, shown in Figure \ref{fig:results_2d}, COS appears to promote strong clumping, which is not observed in the isothermal simulation with $Z = 0.05$ over 1000 orbits. In the former simulation, the Roche density is exceeded after approximately 650 orbits, as indicated by the thick curves. 
The quasi-steady state established in simulations with $Z \geq 0.05$ due to the VSSI persists for several hundred orbits before dust clumps form. We speculate that this clumping is driven by the onset of the SI. This hypothesis is supported by the findings of \citet{johansen2007}, who observed clumping due to the SI in unstratified shearing box simulations for $\tau = 0.1$ and $\epsilon = 1$, conditions similar to the quasi-steady state in our simulations near the mid-plane. However, the present case includes additional turbulence from the VSSI and the effects of vertical gravity on the dust.

In contrast, \citet{li2021} reported strong clumping for $\tau = 0.1$ and $Z = 0.01$ in their stratified shearing box simulations. The absence of similar clumping in our simulations may be attributed to insufficient spatial resolution or differences in simulation geometry. While the finding that COS promotes clumping for $Z=0.05$ is intriguing, it likely holds limited significance, as clumping still occurs in the isothermal simulations at later times ($\sim$1200 ORB, not shown). 
% Furthermore, regions with $Z \gtrsim 0.05$ are expected to be relatively uncommon in typical protoplanetary disks.

%%%%%%%%%%%%%

\subsection{Dust layer thickness and turbulent correlation times}\label{sec:eddy}

%%%%%%%%%%%%%%%%%%%%%%%%%%%%%%%
 \begin{figure}
 \centering 
 	\includegraphics[width=  0.5\textwidth]{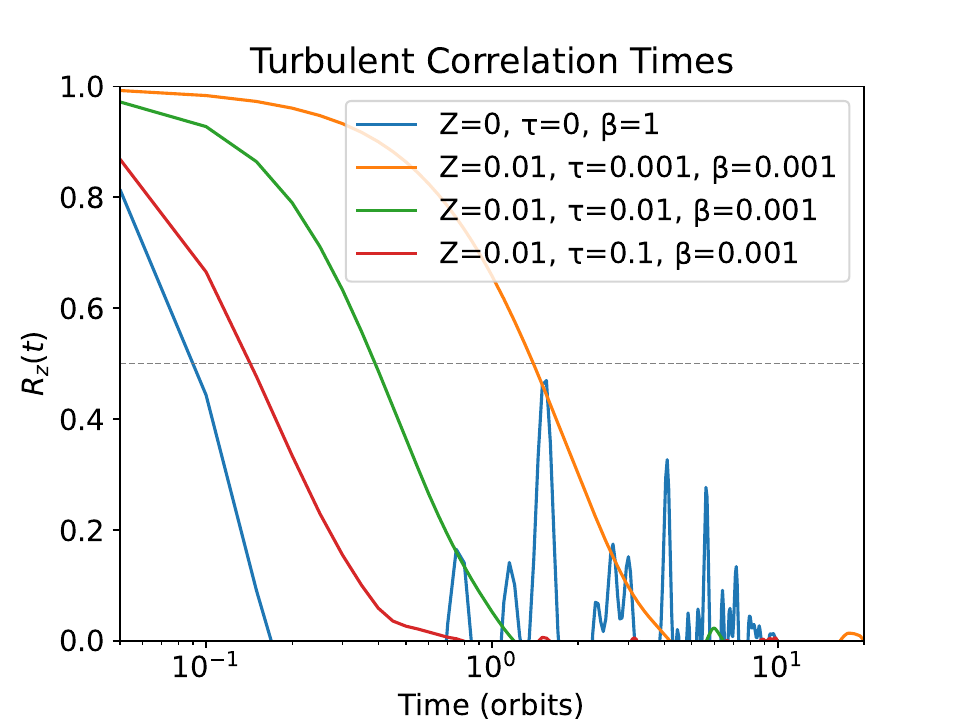}
     \caption{Example auto-correlation functions of the vertical velocity of the dust in isothermal ($\beta=0.001$) dusty 2D simulations with $Z=0.01$, and a non-isothermal ($\beta=1$) pure gas simulation, computed using (\ref{eq:autocorr}) at the mid-plane $z=0$.}
     \label{fig:eddy}
 \end{figure}
%%%%%%%%%%%%%%%%%%%%%%%%%%%%%%%

%%%%%%%%%%%%%%%%%%%%%%%%%%%%%%%%
 \begin{figure*}
 \centering 
 	\includegraphics[width= 0.8 \textwidth]{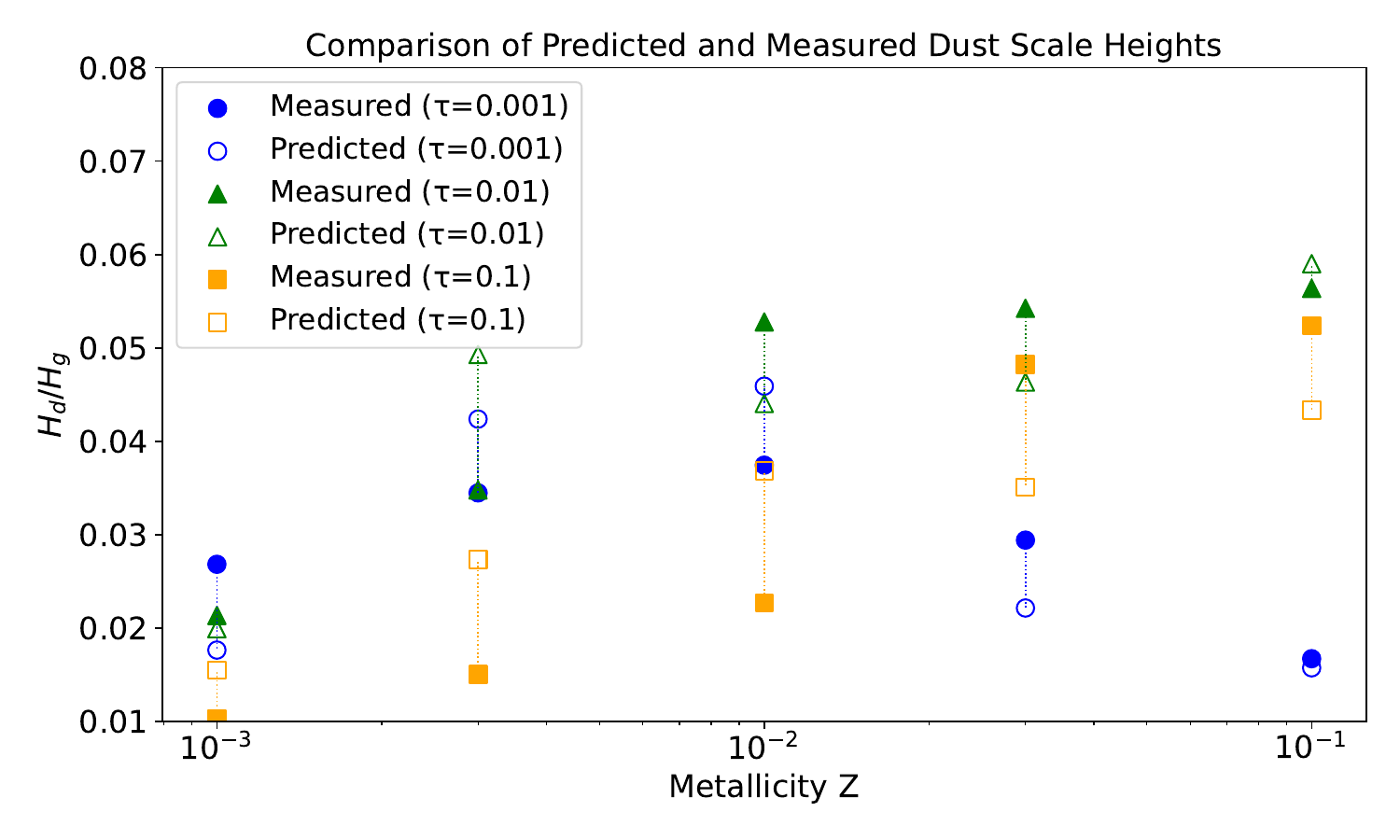}
     \caption{Comparison of measured dust layer thickness as described in Section \ref{sec:eddy}, and averaged over the last 200 orbits in each simulation, with predicted values based on vertical dust profiles obtained from Eq. (\ref{eq:dust_profile}). Compared are 2D simulations with different Stokes number $\tau$ (different symbols) for increasing metallicity $Z$.}
     \label{fig:eddy2}
 \end{figure*}
%%%%%%%%%%%%%%%%%%%

In our simulations presented in Sections \ref{sec:sim_2d} and \ref{sec:sim_3d} we generally found the thickness of the dust layer to exhibit a complex behavior upon variation of the dust parameters $Z$ and $\tau$.
Formally, the dust layer thickness can be expressed as  \citep{dubrulle1995,yl07} (see also Eq. \ref{eq:delta}):
\begin{equation}\label{eq:Hd}
    \frac{H_d}{H_g} = \left( 1 + \frac{c_0^2 \tau}{\langle v_{z}^2 \rangle_t \left(\OmK t_e\right) } \right)^{-1/2},
\end{equation}
assuming a vertical dust diffusion coefficient 
\begin{equation}\label{eq:dustdiff}
D_z \sim \langle v_{z}^2 \rangle_t \, t_e ,
\end{equation}
representing vertical turbulent stirring acting on the dust grains, with the Eddy time (or turbulent correlation time) $t_{e}$. This expression assumes that the quantities $\langle v_{z}^2 \rangle_t$  and $t_e$ are independent of height $z$.

However, since the VSSI is instigated by vertical shear, which is strongly dependent on $z$, and maximal where the dust density drops most rapidly with height, a more sophisticated approach is likely required.
We follow \citet{fromang21} and numerically solve the nonlinear advection diffusion equation for the dust
\begin{equation}\label{eq:dustpde}
\frac{\partial \rho_d}{\partial t} 
- \frac{\partial}{\partial z} \left( z \OmK^2 t_s \rho_d \right) 
= \frac{\partial}{\partial z} \left[ D_z \rho \frac{\partial}{\partial z}  f_d  \right],
\end{equation}
from which the steady-state vertical dust density profile can be obtained via
\begin{equation}\label{eq:dust_profile}
\frac{\partial}{\partial z} \ln f_d  
= -\frac{\OmK^2 t_s}{D_z} z,
\end{equation}

In order to obtain $t_e$ we follow \citet{yang17} and measure the turbulent correlation times related to vertical motions in 2D simulations that have reached a quasi-steady turbulent state, based on the decay of the autocorrelation function
\begin{equation}\label{eq:autocorr}
R_z(t) \equiv \int_{\Delta t}^{} \left[v_{z}(\tau) - \bar{v}_{z}\right] \left[v_{z}(\tau + t) - \bar{v}_{z}\right] d\tau
\end{equation}
with
\begin{equation}
\bar{v}_{z} \equiv \frac{1}{\Delta t} \int_{\Delta t}^{} v_{z}(\tau') d\tau',
\end{equation}
with increasing time $t$. In the above expressions, we choose $\Delta t \sim 20$ orbits. We compute $R_z(t)$ for numerous radial sampling locations throughout the vertical extent of the simulation domain. We then average over all the radial locations to obtain an averaged function $\langle R_z\rangle_{r}(t,z)$.

There is one subtlety involved in the attempt to compare measured dust scaleheights with predicted ones from Eq. (\ref{eq:dust_profile}). We assume that the dust layer maintains a finite thickness through turbulent diffusion within the layer, counteracting settling to the midplane. On top of these internal motions, the dust layer is subjected to a vertically global corrugation mode, similar to the dominant mode seen in pure gas simulations of the VSI. The simple model described by Eq. (\ref{eq:dust_profile}) does not account for corrugation motions.
In order to eliminate the effect of the corrugation motion on the measurement of $H_d$ and $t_e$ we post-process the dust density profile such that we vertically shift the center of mass
\begin{equation}
z_{\text{CMS}}(r) = \frac{\int \rho_d(r,z) z \, dz}{\int \rho_d(r,z) \, dz}
\end{equation}
to the midplane $z=0$. This way, we obtain an approximatelly un-corrugated dustlayer centered on the midplane.
We then also subtract the vertical component of the corrugation velocity from the vertical velocity field at each radius:
\begin{equation}
v_{\text{corr},z}(r) \equiv \frac{\int \rho_d(r,z) v_{z}(r,z) \, dz}{\int \rho_d(r,z) \, dz}
\end{equation}
and then apply the same vertical shift to the resulting vertical velocity field. The measured scaleheight $H_d$ from our simulations is obtained from a Gaussian fit to the un-corrugated dust density profile at each radius, which is then radially averaged.

Figure \ref{fig:eddy} shows the functions $\langle R_i(t)\rangle$ for isothermal example simulations with $Z=0.01$ and Stokes numbers $\tau=0.001$, $\tau=0.01$ and $\tau=0.1$, corresponding to the mid-plane $z=0$. 
In addition, we show the curve for a pure gas simulation with $\beta=1$. The isothermal simulations should reveal the correlation time of VSSI turbulence, whereas the latter simulation should reveal the correlation time of COS turbulence, which we define as the time for which the curves decay by a factor of 0.5. 

We repeat this procedure for all heights so as to obtain $t_e(z)$. In addition, we compute the vertical profile $\langle v_{d,z}^2 \rangle_t$ from the simulation data. Then we integrate (\ref{eq:dust_profile}), and fit a Gaussian to the resulting dust profile to obtain the predicted dust layer thickness $H_d$ and compare these values with the measured values from the simulations. We ran the simulations with $\tau=0.1$, $\tau=0.01$ and $\tau=0.001$ for 1000, 2000, 4000 orbits, respectively, so that all simulations reached a quasi steady state. The results are presented in Figure \ref{fig:eddy2}, showing overall good agreement. However, slight deviations appear to increase with larger values of $\tau$. Whether this trend is coincidental or a result of finite drag effects remains unclear at this stage. As noted in Sections \ref{sec:sim_2d} and \ref{sec:sim_3d}, the thickness of the dust layer exhibits a complex dependence on $Z$ for Stokes numbers $\tau \leq 0.01$. Nonetheless, the results shown here confirm that this behavior aligns well with a simple turbulent settling-diffusion model, which effectively represents turbulent stirring by the VSSI. Finally, we note that $H_d$ was also computed using the simpler model equation (\ref{eq:Hd}), and the overall agreement was only slightly less accurate.
% In section \ref{sec:discuss_axi} we mentioned a sublinear dependence of $max(\epsilon)$ on Stokes number $\tau$ in our axisymmetric simulations, as shown in Figure \ref{fig:vssi_stats}. We now return its discussion. blabla

\subsection{Large-scale vortex formation and evolution}\label{sec:disc_nonaxi}

 \begin{figure}
 \centering 
 	\includegraphics[width=  0.5\textwidth]{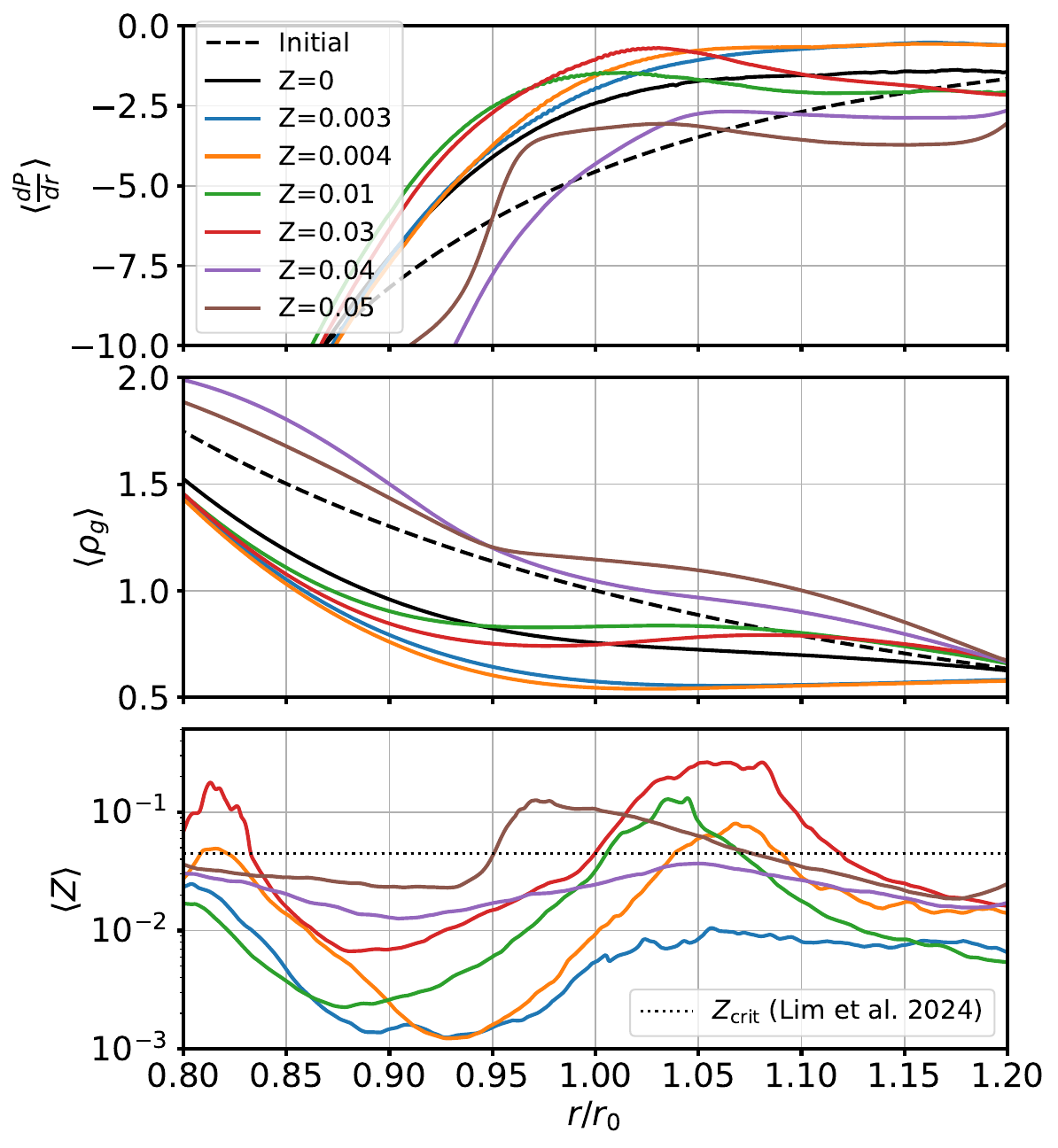}
     \caption{Radial profiles of the azimuthally and time-averaged radial pressure gradient, gas density, and metallicity for simulations with $\tau = 0.1$ and varying initial metallicity Z. The curves for $Z = 0$ represent a pure gas simulation for reference. All colored curves show time-averaged values over the 100 orbits preceding the first occurrence of strong clumping in the simulations. For simulations without strong clumping, the time average was taken over the 100 orbits leading up to the peak value of $\epsilon$. The dotted line in the bottom panel denotes the critical metallicity for strong clumping reported by \citet{lim24}, as explained in Section \ref{sec:disc_clumping}.}
     \label{fig:cos_mitigation_3}
 \end{figure}

In our 3D simulations we find the formation of large-scale vortices, presumably originating from COS-induced zonal flows, which migrate radially inward at rates of $\sim 0.01H$ per orbit (cf. Figure \ref{fig:3d_feedback_1}). However, as pointed out in Section \ref{sec:discuss_axi}, we find zonal flows in all 2D simulations, meaning that not only the COS produces zonal flows. In our 3D simulations the latter are rapidly destroyed at the expense of forming small scale non-axisymmetries or vortices (cf. Figure \ref{fig:vssi_cos}). It is assumed that large-scale vortices form through merging of multiple small-scale vortices \citep{manger2018,raettig2021,lehmann2022}.

For a Stokes number $\tau=0.1$ large-scale vortices are found to form for $Z\lesssim 0.03$ (cf. the discussion of Figure \ref{fig:cos_mitigation_1}).
For larger metallicity, these are absent, although small-scale short lived vortices still form. Interestingly, this suppression threshold is close to the value reported by \citet{lehmann2022} for suppression of large vortices due to the VSI at $\tau = 0.01$.
The absence of large-scale vortices could be explained if the small-scale vortices are too short-lived, or if their mutual merging becomes inefficient at sufficiently large $Z$. 
The ability to merge should be mostly tied to the global radial gas density profile \citep{paardekooper2010,lehmann2022}. 

 Figure \ref{fig:cos_mitigation_3} shows radial profiles of various quantities in the nonlinear saturated state of 3D simulations with $\tau=0.1$ and increasing $Z$. Notably,  we see a significant change in the radial gas density profile for $Z\geq0.04$, compared to smaller $Z$. The density profile becomes less flattened for $r/r_0\gtrsim 1$, which is where large-scale vortices which undergo strong clumping form in the simulations with smaller $Z$. A steeper gas density profile should make it harder for vortices to merge, based on the findings of \citet{lehmann2022}. The different evolution of the radial gas density profile for smaller $Z$ arises from the stronger radial angular momentum transport in these simulations, as evident in Figure \ref{fig:cos_mitigation_1} (top panel). Nevertheless, the significance of this effect remains somewhat speculative, and a more rigorous analysis of vortex migration is beyond the scope of this paper. In particular, it does not explain the initially reduced angular momentum transport observed for $Z\geq 0.04$.

 \begin{figure}
 \centering 
 	\includegraphics[width=  0.5\textwidth]{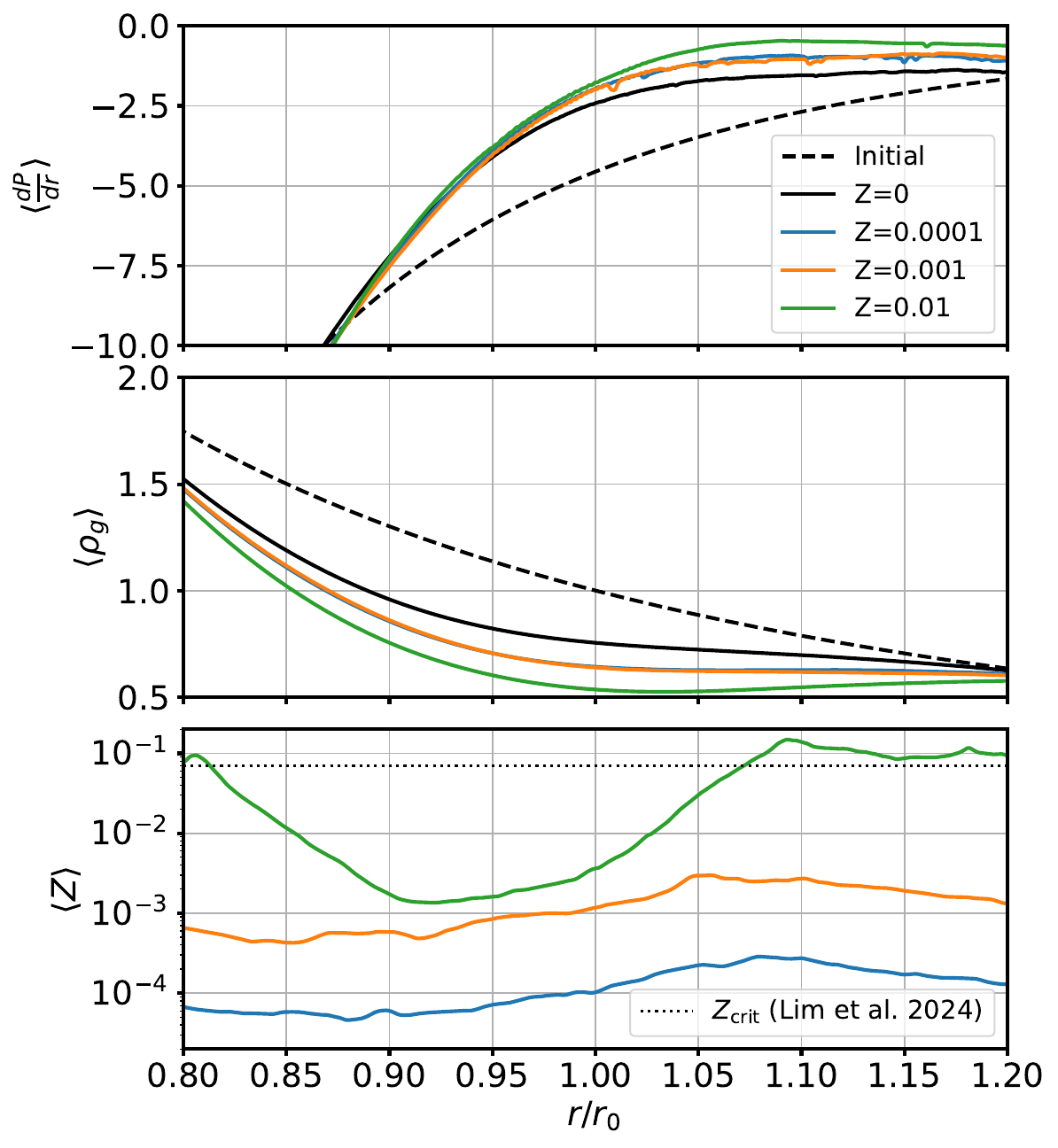}
     \caption{Same as Figure \ref{fig:cos_mitigation_3}, but for simulations with a Stokes number $\tau=0.05$.}
     \label{fig:raettig_comp_2}
 \end{figure}

In addition, it is possible that the life time of vortices is reduced by the onset of drag instabilities (Section \ref{sec:instabilities}) within them. Since these instabilities are expected to instigate velocity perturbations throughout the entire vertical gas column (which we confirmed for select vortices), this can result in a disruption of the entire vortex column, and not only the mid-plane region where the dust density is large.
This interpretation is supported by our result that large-scale vortices persist in simulations with $\tau = 0.01$ up to metallicities of $Z = 0.1$. Note also that a larger value of $\tau$ implies an overall thinner dust layer. Therefore, the midplane region of vortices that could be disrupted due to dust-loading should be even smaller if $\tau$ is larger. However, drag instabilities, such as the VSSI and SI, excite more vigorous turbulent vertical motions with increasing $\tau$ (cf. the isothermal curves in Figure \ref{fig:results_2d}), such that vortices are expected get more strongly disrupted.
Similarly, the increased onset of drag instabilities might contribute to the observed transition of the shape of dust structures from non-axisymmetric to axisymmetric as discussed in Section \ref{sec:duststructures}.

Finally, it can be expected that the COS is somewhat weakened in simulations with $Z\geq 0.04$ due to dust loading. For these simulations we find midplane dust-to-gas ratios of $\epsilon\gtrsim 1$. According to the local analysis of \citet{lehmann2023} linear COS growth rates are proportional to a factor $1/(1+\epsilon)$, implying a significant reduction. It is unclear though, how vertical dust stratification and nonlinearity affect this result.

\subsection{Comparison with existing 3D shearing box simulations of the COS}

The results of our 3D simulations can, to some extent, be compared with the shearing-box simulations of \citet{raettig2021}. 
% Figure \ref{fig:raettig_comp_1} shows the time evolution of $\epsilon_{\text{max}}$, as well as the mass fraction of the disk where $\epsilon$ exceeds specific thresholds, for
To this end, we performed simulations with $\tau = 0.05$ and $Z = 10^{-2}$, $Z = 10^{-3}$, and $Z = 10^{-4}$. These same parameters were also considered by \citet{raettig2021}, who reported $\epsilon_{\text{max}} > 1$ in all three cases. More precisely, they found $\epsilon_{\text{max}} \gtrsim 1$, $\epsilon_{\text{max}} \gtrsim 10$, and $\epsilon_{\text{max}} \gtrsim 100$ for the respective values of $Z$. In our simulations, the values of $\epsilon_{\text{max}}$ are approximately 10 times smaller for $Z = 10^{-3}$ and $Z = 10^{-4}$. In contrast, for $Z = 10^{-2}$, we find similar or even higher values of $\epsilon_{\text{max}}$. Notably, strong clumping occurs only in the simulation with $Z = 10^{-2}$.

Possible explanations for this discrepancy include differences in numerical methods (dusty fluid vs. Lagrangian super-particles) and the geometrical setup (radially global vs. radially local with periodic boundaries). Regarding the latter, Figure \ref{fig:raettig_comp_2} illustrates for our simulations with $\tau=0.05$ the evolution of radial profiles for various quantities, similar to Figure \ref{fig:cos_mitigation_3}. Significant changes occur in the radial profiles across all three simulations, with the radial pressure gradient directly influencing the dust distribution within the simulation domain. 
%\mkl{is it generic that feedback reduces |dP/dr|?} \ml{(RE: I don't think so. For smaller $Z\lesssim 0.01$ feedback increases changes on the radial disc structure due to vortices (VSSI vorticity driving), including dP/dr, whereas for larger $Z\gtrsim 0.04$ it reduces these changes (suppression of large vortices), see Figure 15.)}

In all three cases, a large-scale zonal flow emerges near $r / r_0 = 1.1$, which subsequently spawns vortices. As discussed in the previous section, these vortices likely form through the merging of smaller-scale vortices and then migrate radially inward. Notably, the carved pressure bump is most pronounced in the simulation with $Z = 0.01$, as vortex formation begins significantly earlier in this case. We assume that this early onset is driven by a more vigorous VSSI, which immediately generates small-scale, non-axisymmetric flow structures (cf. Figure \ref{fig:vssi_cos}). Consequently, dust accumulates in the outer regions of the simulation domain, leading to a concentration of dust in large-scale vortices. In fact, in this simulation, $Z$ increases by more than an order of magnitude in the outer regions, whereas only order-unity increases are observed in the other simulations. Such global effects are not captured in the radially local simulations of \citet{raettig2021}.

That all being said, the agreement between our results and those of \citet{raettig2021} for $Z = 0.01$ may be a fortunate outcome of global effects, which are stronger in this case. For $Z = 10^{-3}$ and $Z = 10^{-4}$, these effects appear too weak to drive comparable results. 

Additionally, while \citet{raettig2021} report inward migration of dust-laden vortices, it is unclear how this result compares to ours, as we simulate the coupled evolution of dust and gas from the beginning, with significantly longer simulation times. In contrast, \citet{raettig2021} add dust only after the COS has saturated into large-scale vortices. At the same time, their numerical scheme is constructed in a way that in the absence of dust, no radial background pressure gradient is present\footnote{This is evident from their Equations (2) and (9) if $\rho_d \to 0$}, which is required for vortex migration \citep{paardekooper2010}.

Furthermore, periodic boundaries in their simulations imply that vortices that cross the inner boundary will reappear in the domain, enabling them to grow further.
This might imply that the vortices in their simulations are larger on average, allowing them to concentrate dust more effectively, especially when multiple dust-laden vortices merge.

As discussed in Section \ref{sec:feedback}, we observed that the vortex shown in Figure \ref{fig:3d_feedback_1} (located around $r / r_0 = 1$ at 800 orbits) undergoes strong clumping with $\rho > \rho_{\text{Roche}}$ for approximately 50 orbits before being completely destroyed. Prior to its dispersal, a dense dust clump is expelled from the vortex. It remains unclear whether dust feedback is responsible for the vortex’s demise. In contrast, \citet{raettig2021} concluded that vortices in their simulations do not disperse away from the midplane, even at high $\epsilon$ values. However, their simulations span only about 20 orbits, which may not be long enough to capture the eventual destruction of vortices.

\subsection{Critical Metallicity for strong dust clumping}\label{sec:disc_clumping}

It is interesting to note that our critical initial metallicity $Z_{\text{crit}} \approx 0.004$ for strong clumping at $\tau = 0.1$ (see Figure \ref{fig:cos_clumping}) closely matches the value reported by \citet{li2021}, who performed 3D shearing box simulations of the SI, and who found $Z_{\text{crit}} = 0.006$. Furthermore, we observe clumping for $\tau \gtrsim 0.04$ at initial solar metallicity ($Z = 0.01$), whereas \citet{li2021} reported strong clumping for $\tau = 0.02$. On the other hand, for $\tau=0.01$, \citet{li2021} find $Z_{crit}=0.02$, whereas we don't find strong clumping even for $Z=0.1$.

More recently, \citet{lim24} extended the study of \citet{li2021} by incorporating the effects of self-gravity and external turbulence on the dust particles. Their simulations included forced isotropic turbulence with $\alpha = 10^{-4}\text{–}10^{-3}$. They found significantly higher critical metallicities for strong clumping compared to \citet{li2021}. Using $\alpha=10^{-3}$, for $\tau = 0.01$, they reported strong clumping for $Z \geq 0.25$, which is consistent with our result of no clumping for $Z \leq 0.1$. However, for $\tau = 0.1$, they observed strong clumping only for $Z \geq 0.045$, approximately an order of magnitude higher than our initial value. This discrepancy arises even though turbulence levels in our simulations yield $\alpha_r \sim 10^{-3}\text{–}10^{-2}$, and even though our setup omits self-gravity, which would be expected to lower the critical metallicity for clumping. The inclusion of self-gravity in our simulations would likely further increase the difference between our results and those of \citet{lim24}.

In the discussion so far, we have deliberately used the term “initial metallicity”. The discrepancies with \citet{lim24} can be resolved if we consider a more appropriate measure of metallicity. In our simulations, we observe significant evolution in the radial disk structure, accompanied by changes in the radial profile of metallicity. Returning to Figure \ref{fig:cos_mitigation_3}, the bottom panel shows time- and azimuth-averaged metallicity profiles, where the time average was taken over the last 100 orbits before strong clumping occurred. In simulations without strong clumping, the averaging was performed over the 100 orbits preceding the peak value of $\epsilon$. The dotted horizontal line represents the critical metallicity, $Z_{\text{crit}}$, reported by \citet{lim24} for $\tau = 0.1$ and turbulence level $\alpha = 10^{-3}$. Interestingly, all simulations that exhibited strong clumping exceeded this critical metallicity when using the described averaged metallicity, whereas those without strong clumping remained below it.

The same pattern is observed in the three simulations shown in Figure \ref{fig:raettig_comp_2}. In this case, the dotted line in the bottom panel represents the critical metallicity reported by \citet{lim24} for $\tau = 0.05$. Again, the simulation with an initial $Z = 0.01$ exhibited strong clumping, while the others did not. However, this agreement does not hold universally across all values of $\tau$. In simulations with $\tau = 0.01$, we do not observe strong clumping even for an initial metallicity of $Z = 0.1$, despite global effects raising the metallicity to exceed the critical value reported by \citet{lim24} ($Z_{\text{crit}} = 0.25$). It is important to note that our grid resolution is significantly lower than that of \citet{lim24}, which may inhibit clumping in this case. Additionally, the turbulence level applied in \citet{lim24} ($\alpha = 10^{-3}$) may slightly underestimate the effects of turbulence in our simulations, where we typically find $\alpha$-values that are several times larger.

Whether the good agreement in the critical metallicity for $\tau = 0.1$, as well as the critical Stokes number at solar initial metallicity, with \citet{li2021} is a fortunate coincidence or reflects a deeper underlying connection remains speculative, particularly given the significant differences in the simulation setups. The absence of dust diffusion in their simulations, combined with the influence of global effects in ours, could together contribute to this agreement.

\subsection{Implications for planetesimal formation}

Our results demonstrate that the COS has significant potential to actively promote planetesimal formation through the generation of large-scale vortices, without requiring pre-existing density structures, such as gaps or pressure bumps, as long as the condition $N_r^2<0$ is met.

For suitable dust parameters, these vortices persist long enough to efficiently concentrate dust into dense clumps.

We observe a strong dependence of the clumping tendency on the Stokes number $\tau$. Specifically, for $\tau = 0.01$, strong clumping does not occur at initial metallicities $Z \lesssim 0.1$ (and potentially even higher).
In this regime, dust is tightly coupled to the gas and remains relatively well-mixed within the vortex flow, rather than concentrating at the vortex center \citep[e.g.,][]{meheut2012}. Consequently, $\epsilon$ stays small, limiting the backreaction of dust onto the gas. As a result, large-scale vortices persist as stable, long-lived structures.
In contrast, for $\tau = 0.1$, strong clumping is observed at sub-solar initial metallicities, specifically for $Z \gtrsim 0.004$.

For larger Stokes numbers ($\tau \gtrsim 0.04$), dust grains can partially decouple from the gas and concentrate more effectively. This stronger clumping enhances the dust-to-gas ratio in the vortex core, which can disrupt the vortex and shorten its lifespan, as discussed in Section \ref{sec:disc_nonaxi}. Hence, while large-scale vortices do form for $\tau \geq 0.04$ (at least at solar metallicity $Z=0.01$), their lifetime is reduced by strong dust feedback.
Under these conditions, vortices in our simulations are found to be significantly weaker and more spatially extended (cf. Figure \ref{fig:3d_feedback_1}).

\subsection{Implications for observations}
It is noteworthy that the outer of two axisymmetric dust rings observed in HD 163296, located at 100 AU from the central star, as analyzed by \citet{doi2023} using ALMA bands 4 and 6, shows promising compatibility with the results of our 3D simulations. Specifically, these authors inferred a Stokes number $\tau > 0.1$, a turbulent parameter $\alpha_r > 0.005$, and a dust-to-gas scale height ratio $H_d/H_g < 0.08$, with a local disk aspect ratio $h \sim 0.065$ from ALMA band 4 observations. These values align well with the results of our simulations with $\tau = 0.1$ (and $h_0 = 0.1$), as shown in Figure \ref{fig:cos_mitigation_1}. Furthermore, since the authors also report a ratio $\alpha_z/\alpha_r < 0.19$ for the outer ring, it is unlikely that the VSI is the dominant source of turbulence in this region of the disk. For instance, \citet{lehmann2022} found $\alpha_z/\alpha_r \sim 2-3$ for the VSI \footnote{this can be seen from their Figures 6 and 11}. By contrast, in \PaperI, we found that $\alpha_z$ is substantially smaller than $\alpha_r$ for the COS, highlighting a key distinction in the turbulent characteristics of the COS and VSI. \footnote{Note that this result may differ if vertical stratification of the gas is included.} 
%\mkl{not clear why their observation is consistent with our results?} \ml{(RE: I don't know of any other plausible instability that can produce such $\alpha$ values, if we exclude MRI and VSI.)}

% \mkl{the following isn't too relevant to this paper (unless we demonstrated that alphaz/alphar is not anisotropic for COS)}
For the inner ring at 67 AU, by contrast, \citet{doi2023} identified a pronounced anisotropy, with $\alpha_z/\alpha_r \gg 1$. This feature is inconsistent with COS-driven turbulence, as noted earlier, but could potentially be attributed to the presence of the VSI. The coexistence of VSI at inner radii and its absence at larger radii is plausible if the effects of dust coagulation and fragmentation on gas cooling are considered, as demonstrated by \citet{pfeil2024}. However, further observations, such as measurements of local radial gas density and temperature gradients, are needed to confirm whether COS could be active in this region and whether it contributes to the observed turbulence characteristics.
% %
% The finding that the observed rings are axisymmetric suggests, based on our results, a super-solar metallicity. However, in our simulations with $\tau=0.1$ and $Z = 0.01$, vortices are not very long-lived \mkl{do you mean the gas vortex? but observations are in dust}; they eventually dissolve into a turbulent dust ring before new vortices are regenerated. For $Z \gtrsim 0.02$, vortices become significantly less long-lived. Whether the observed rings are resolved well enough to detect vortices, or whether the observations sampled the rings during a state without vortices, remains unclear.

%\mkl{i think the discussion on observations needs a little work. are you saying the doi23 outer ring can result from COS because 1) our results indicate that COS creates pressure bumps to trap dust, but if Z is large enough it will remain as a ring and 2) COS turbulence is isotropic (is this the case for doi23's outer ring?)}
%\ml{(RE: Since we don't have relevant data on disc structure,  I don't think there is more we can do here apart from speculating. 1) yes, possibly. 2) COS turbulence is not isotropic, rather anisotropic with alphar $\gg$ alphaz (at least in 3D), therefore incompatible with the inner ring, but still possibly compatible with the outer ring.)}

\subsection{Caveats and Outlook}

As in \hyperlinkcite{lehmann2024}{Paper I}, our simulations employ a simplified optically thin cooling law characterized by a single parameter, $\beta$. However, recent studies suggest that cooling times conducive to the COS are likely associated with optically thick conditions near the disk mid-plane (e.g., \citealt{pfeil2019}). Future work should explore more sophisticated cooling models, incorporating thermal diffusion or radiative transfer, to better capture the thermal dynamics. Even within the simple optically thin cooling regime, variations in the cooling time significantly impact the strength and lifetimes of vortices (\hyperlinkcite{lehmann2024}{Paper I}), a factor not fully explored in the current study.

We note that while introducing more complex cooling models would improve the physical realism of the simulations, the current setup already presents substantial challenges. It involves at least four interacting instabilities—the COS, VSSI, SI, and elliptic instability—which collectively complicate the interpretation of the results. Incorporating more sophisticated cooling physics would add further complexity, making it even more challenging to disentangle the contributions of individual instabilities.

Expanding the radial domain and increasing grid resolution would further enhance the accuracy of the simulations, particularly by reducing boundary effects and improving the resolution of turbulence and clumping within vortices. These enhancements remain a technical challenge for now but are a priority for future studies as advancements in computational resources allow.

Another open question is the precise mechanism responsible for the suppression of large-scale vortices at sufficiently high metallicity ($Z \geq 0.04$ for $\tau=0.1$ in our simulations). As discussed in Section \ref{sec:disc_nonaxi}, this suppression could be linked to inefficient vortex mergers due to modifications in the radial gas density profile or to increased dust loading, which may reduce COS growth rates. Additionally, drag instabilities such as the VSSI and SI could generate vertical velocity perturbations that disrupt the vortex column. However, the relative importance of these effects remains uncertain, and future studies should aim to systematically disentangle their roles.

Eventually, fully global 3D models with realistic disk structures, such as density bumps and jumps as considered in \hyperlinkcite{lehmann2024}{Paper I}, should be investigated. However, achieving this level of complexity is far beyond the current feasibility at the desired resolution. In a follow-up study, we plan to examine the influence of nonlinear COS on an embedded, migrating planet within a protoplanetary disk.

\bigskip
%\nolinenumbers  % Turn off line numbers

This work is supported by the National Science and Technology
Council (grants 112-2112-M-001-064-, 113-2112-M-001-036-
112-2124-M-002-003-, 113-2124-M-002-003-) and an Academia
Sinica Career Development Award (AS-CDA-110-M06). Simulations were performed on the \emph{Kawas} cluster at ASIAA and the \emph{Taiwania-2} cluster at the National Center for High-performance Computing (NCHC). We thank NCHC for
providing computational and storage resources. 

%\relax  % Ensure LaTeX does not re-enable them

% \newpage

%  % \appendix

% \section{Turbulend Eddy times}\label{app:eddy}

\end{document}